\documentclass[fleqn,usenatbib]{mnras}

\usepackage{newtxtext,newtxmath}

\usepackage[T1]{fontenc}

\DeclareRobustCommand{\VAN}[3]{#2}
\let\VANthebibliography\thebibliography
\def\thebibliography{\DeclareRobustCommand{\VAN}[3]{##3}\VANthebibliography}

\usepackage{graphicx}	
\usepackage{amsmath}	
\usepackage{amssymb}	
\usepackage{bm}
\usepackage{caption}
\usepackage{multicol}
\usepackage{hyperref}
\usepackage{booktabs}
\usepackage{ulem}
\usepackage{mathtools}
\usepackage{lipsum} 

\def\eqref#1{equation~(\ref{#1})}

\newcommand {\Myr}{\,{\rm Myr}}
\newcommand {\Gyr}{\,{\rm Gyr}}

\newcommand {\kpc}{\,{\rm kpc}}
\newcommand {\kms}{\,{\rm km}\,{\rm s}^{-1}}

\newcommand {\Msun}{\,{\rm M}_\odot}

\newcommand {\vc}{v_{\rm c}}

\newcommand {\drm}{\mathrm{d}}

\newcommand {\vx}{{\bm x}}
\newcommand {\vvel}{{\bm v}}

\newcommand {\Lz}{L_z}

\newcommand {\phib}{\varphi_{\rm b}}

\newcommand {\rCR}{r_{\rm CR}}
\newcommand {\rb}{r_{\rm b}}
\newcommand {\vJ}{{\bm J}}

\newcommand {\Jr}{J_r}

\newcommand {\vJf}{{\bm J}_{\rm f}}
\newcommand {\Jfo}{J_{{\rm f}_1}}
\newcommand {\Jft}{J_{{\rm f}_2}}
\newcommand {\Js}{J_{\rm s}}

\newcommand {\Jsres}{J_{\rm s, res}}

\newcommand {\dJsres}{\dot{J}_{\rm s, res}}

\newcommand {\dJres}{\delta J_{\rm res}}
\newcommand {\ddJres}{\delta \dot{J}_{\rm res}}
\newcommand {\Jl}{J_\ell}

\newcommand {\Jlsep}{J_{\ell,{\rm sep}}}

\newcommand {\vtheta}{{\bm \theta}}

\newcommand {\thetar}{\theta_r}

\newcommand {\thetaphi}{\theta_\varphi}
\newcommand {\thetapsi}{\theta_\psi}
\newcommand {\thetas}{\theta_{\rm s}}
\newcommand {\dthetas}{\dot{\theta}_{\rm s}}

\newcommand {\thetafo}{\theta_{{\rm f}_1}}
\newcommand {\thetaft}{\theta_{{\rm f}_2}}

\newcommand {\thetares}{\theta_{\rm res}}
\newcommand {\thetasres}{\theta_{\rm s,res}}
\newcommand {\thetasep}{\theta_{\rm sep}}
\newcommand {\thetasepm}{\theta_{\rm sep}^{-}}
\newcommand {\thetasepp}{\theta_{\rm sep}^{+}}

\newcommand {\thetal}{\theta_\ell}

\newcommand {\vthetaf}{{\bm \theta}_{\rm f}}
\newcommand {\Omegap}{\Omega_{\rm p}}
\newcommand {\Omegapo}{\Omega_{\rm p0}}
\newcommand {\Omegapi}{\Omega_{\rm p1}}
\newcommand {\dOmegap}{\dot{\Omega}_{\rm p}}

\newcommand {\vOmega}{\mathbf \Omega}

\newcommand {\Omegas}{\Omega_s}

\newcommand {\vn}{{\bm n}}
\newcommand {\vN}{{\bm N}}
\newcommand {\vk}{{\bm k}}

\newcommand {\nr}{n_r}
\newcommand {\npsi}{n_\psi}
\newcommand {\nphi}{n_\varphi}

\newcommand {\Nr}{N_r}
\newcommand {\Npsi}{N_\psi}
\newcommand {\Nphi}{N_\varphi}
\newcommand {\ks}{k_{\rm s}}

\newcommand {\kfo}{k_{{\rm f}_1}}
\newcommand {\kft}{k_{{\rm f}_2}}

\newcommand {\Phib}{\Phi_{\rm b}}

\newcommand {\Phin}{\Phi_{\vn}}

\newcommand {\PhiN}{\Phi_{\vN}}

\newcommand {\Psik}{\Psi_{\vk}}

\newcommand {\whPhin}{\widehat{\Phi}_{\vn}}

\newcommand {\bH}{\bar{H}}
\newcommand {\hH}{\widehat{H}}

\newcommand {\e}{\mathrm{e}}
\newcommand {\sgn}{\mathrm{sgn}}

\newcommand {\Td}{T_{\rm d}}

\newcommand {\Tl}{T_\ell}

\newcommand {\rs}{r_{\rm s}}

\newcommand {\pd}{{\partial}}

\newcommand {\taufb}{\tau_{\rm fb}}

\newcommand {\taul}{\tau_{\ell}}
\newcommand {\tauc}{\tau_{\rm c}}
\newcommand {\htau}{\hat{\tau}}
\newcommand {\htaufb}{\hat{\tau}_{\rm fb}}
\newcommand {\htaul}{\hat{\tau}_{\ell}}
\newcommand {\htauc}{\hat{\tau}_{\rm c}}
\newcommand {\Sl}{S_{\ell}}
\newcommand {\fl}{f_{\ell}}

\newcommand {\fmix}{\tilde{f}}
\newcommand {\flmix}{\tilde{f}_{\ell}}
\newcommand {\fcmix}{\tilde{f}_{\rm c}}

\newcommand {\Bl}{B_{\ell}}
\newcommand {\dLz}{\dot{L}_z}
\newcommand {\dDelta}{\dot{\Delta}}
\newcommand {\Ifb}{I_{\rm fb}}

\newcommand {\Mb}{M_{\rm b}}
\newcommand {\Irig}{I_{\rm rigid}}

\title[]{Dynamical friction and feedback on galactic bars in the general fast-slow regime}

\author[R. Chiba]{
Rimpei Chiba$^{1,2,3}$\thanks{E-mail: rchiba@cita.utoronto.ca}
\\
$^{1}$Canadian Institute for Theoretical Astrophysics, University of Toronto, 60 St. George Street, Toronto, ON M5S 3H8, Canada\\
$^{2}$Mullard Space Science Laboratory, University College London, Holmbury St. Mary, Dorking, Surrey, RH5 6NT, UK\\
$^{3}$Rudolf Peierls Centre for Theoretical Physics, University of Oxford, Parks Road, Oxford, OX1 3PU, UK
}

\date{Accepted XXX. Received YYY; in original form ZZZ}

\pubyear{2023}

\begin{document}
\label{firstpage}
\pagerange{\pageref{firstpage}--\pageref{lastpage}}
\maketitle

\begin{abstract}

Current theories of dynamical friction on galactic bars are based either on linear perturbation theory, which is valid only in the \textit{fast limit} where the bar changes its pattern speed rapidly, or on adiabatic theory, which is applicable only in the \textit{slow limit} where the bar's pattern speed is near-constant. In this paper, we study dynamical friction on galactic bars spinning down at an arbitrary speed, seamlessly connecting the fast and slow limits. We treat the bar-halo interaction as a restricted $N$-body problem and solve the collisionless Boltzmann equation using the fast-angle-averaged Hamiltonian. The phase-space distribution and density wakes predicted by our averaged model are in excellent agreement with full 3D simulations. In the slow regime where resonant trapping occurs, we show that, in addition to the frictional torque, angular momentum is transferred directly due to the migration of the trapped phase-space: trapped orbits comoving with the resonance typically gain angular momentum, while untrapped orbits leaping over the trapped island lose angular momentum. Due to the negative gradient in the distribution function, gainers typically outnumber the losers, resulting in a net negative torque on the perturber. Part of this torque due to the untrapped orbits was already identified by Tremaine \& Weinberg who named the phenomenon \textit{dynamical feedback}. Here, we derive the complete formula for dynamical feedback, accounting for both trapped and untrapped orbits. Using our revised formula, we show that dynamical feedback can account for up to $30\%$ of the total torque on the Milky Way's bar.

\end{abstract}




\defcitealias{lynden1972generating}{LBK72}
\defcitealias{Tremaine1984Dynamical}{TW84}
\defcitealias{Weinberg2004Timedependent}{W04}

\section{Introduction}
\label{sec:introduction}

Dynamical friction is a process by which a heavy object moving in space gravitationally transfers net momentum to the surrounding matter \citep{Chandrasekhar1943DynamicalFriction}. It is one of the key processes that drives the secular evolution of galaxies, most notably the spin-down of galactic bars \citep[e.g.][]{weinberg1985evolution,hernquist1992bar,Athanassoula1996Evolution} and the orbital decay of satellite galaxies \citep[e.g.][]{Lin1983Numerical,White1983Simulations,Weinberg1986OrbitalDecay}. Dynamical friction is important for probing dark matter, as it depends not only on the existence of dark matter \citep[e.g.][]{Tiret2007MOND,debattista2000constraints,Ghafourian2020Modified}, but also on its kinematic distribution \cite[e.g.][]{athanassoula2003determines,Long2014Secular,Sellwood2016BarInstability} and even on its particle nature \citep[e.g.][]{Hui2017Ultralight,Lancaster2020DynamicalFrictionFDM,Wang2022Dynamical}.

The original study of dynamical friction by \cite{Chandrasekhar1943DynamicalFriction} considered the drag force on a point-mass perturber moving through an infinite homogeneous background of particles. Under the assumption of homogeneity (i.e. a flat background potential), field particles undergo a single hyperbolic encounter with the perturber. However, in realistic stellar systems with a finite inhomogeneous distribution, most particles are gravitationally bound to the system, meaning that they encounter the perturber periodically. Due to this periodic interaction, the exchange of momentum between the particle and the perturber becomes most effective when the two are in orbital resonance.

\citet[][hereafter \citetalias{Tremaine1984Dynamical}]{Tremaine1984Dynamical} were the first to describe dynamical friction in terms of resonances. Using perturbation theory, they derived an analytic expression for dynamical friction in spherical systems which states that friction arises exclusively from resonant orbits. Their dynamical friction formula is referred to as the LBK formula, named after \cite{lynden1972generating} who derived a similar equation in the context of disk dynamics. \citetalias{Tremaine1984Dynamical} further showed that the quality of dynamical friction changes depending on the speed at which the resonance sweeps the phase space. They classified the evolution based on the dimensionless \textit{speed} of the resonance $s$, where they called 
$0 \leq s < 1$ `slow' and $1 \leq s$ `fast'.

Our current understanding on dynamical friction is limited to either of the two extreme evolutions: the \textit{fast limit} ($s \gg 1$) or the \textit{slow limit} ($s \ll 1$). Fast limit refers to a perturber rapidly changing its frequency such that the resonances sweep the orbits before any nonlinearity can develop. In this limit, a time-dependent linear perturbation theory\footnote{For a time shorter than the libration period, linear theory works regardless of the speed of the resonance \citep[e.g.][]{Magorrian2021PeriodicCube} (see also Section \ref{sec:comparison_linear_theory}).
Since the libration period scales as the inverse square root of the perturbation amplitude, linear theory holds if the perturbation is weak \citep[e.g., Poisson shot noise,][]{Fouvry2015Secular} or short lived \citep[e.g., transient spiral arms,][]{Carlberg1985Dynamical}.} gives a decent approximation \citep{Weinberg2004Timedependent,Banik2021SelfConsistent}. It is worth noting that the time-dependent linear-torque formula reduces to the classical LBK formula if one assumes that (i) an infinite time has passed since the perturber emerged (the \textit{time-asymptotic limit}) and (ii) the perturber rotates at a fixed pattern speed (the \textit{slow limit})\footnote{Some authors refer to the time-asymptotic limit as the \textit{adiabatic approximation} and the slow limit as the \textit{secular approximation} \citep{Banik2021SelfConsistent,Kaur2022MNRASDensity}}. The linear approximation, however, breaks down precisely at the intersection of these two limits \citep[e.g.][]{Chiba2022Oscillating}. Thus the LBK formula is at best qualitative. The structure of the density wakes corresponding to the LBK formula was studied in detail by \cite{Kaur2022MNRASDensity}. 

The slow limit is the opposite extreme where the perturber evolves at a near-constant frequency. In this limit, the orbital response near resonances grows nonlinearly and saturates due to resonant trapping. Trapping is a nonlinear effect which cannot be treated with linear theory. In the slow limit, however, the change in the perturbing frequency is assumed to be slow enough that the evolution of the trapped orbits can be modelled by identifying their adiabatic invariants by the method of averaging \citep[e.g.][]{Neishtadt1975Passage,Henrard1982Capture,Sridhar1996Adiabatic}. \cite{Chiba2022Oscillating} formulated dynamical friction on galactic bars in this limit and showed that the phase mixing of the trapped orbits gives rise to an oscillating friction that damps over several libration periods. \cite{Banik2022Nonperturbative} proposed a similar mechanism to explain the oscillation and subsequent stalling of sinking satellites at the core of galaxies \citep[e.g.][]{Read2006MNRASConstantDensityCore}.

Unfortunately, neither the fast nor slow limit is appropriate for describing the evolution of real galaxies. The perturber inevitably changes its frequency in response to dynamical friction, yet the rate of change is not always fast enough to justify the linear approximation. For example, both $N$-body simulations \citep[e.g.][]{Petersen2016DarkMatterTrapping,Halle2018Radial} and observations \citep[e.g.][]{Monari2019signatures,Binney2020Trapped,Chiba2021TreeRing} indicate that important evolutions of barred galaxies occur in the general slow regime where substantial amount of stars and dark matter are captured into resonance. To complicate things further, some orbits may evolve in the fast regime while others evolve in the slow regime, since $s$ depends not only on the rate of change of the pattern speed but also on the libration frequency of individual orbits, which varies with resonances and with the position along each resonance.

The only study that has addressed dynamical friction in the general fast-slow regime is \citetalias{Tremaine1984Dynamical}. By solving the nonlinear orbital motions, they formulated the net change in the angular momentum of orbits swept by resonances in the time-asymptotic limit. They found that untrapped orbits swept by a slowly moving resonance provide a torque that directly depends on the rate of change of the pattern speed and therefore modifies the effective moment of inertia of the perturber. \citetalias{Tremaine1984Dynamical} termed this torque \textit{dynamical feedback} to distinguish it from dynamical friction. There are, however, two issues in \citetalias{Tremaine1984Dynamical}'s formalism: First, because they take the time-asymptotic limit, their theory fails to capture the transient behaviour of dynamical friction which lasts for several libration periods and can therefore have a significant impact on the perturber's evolution. Secondly, they did not take into account (though discussed) the dynamical feedback exerted by trapped orbits. In view of the conservation of phase-space volume, the loss (gain) of angular momentum by untrapped orbits leaping over the resonance and the gain (loss) by the trapped orbits dragged by the resonance occur simultaneously and should be considered together.

In this paper, we address the shortcomings of \citetalias{Tremaine1984Dynamical} in the context of bar-halo interaction. First, we study dynamical friction and feedback using a semi-numerical approach, i.e., we numerically solve the collisionless Boltzmann equation using a Hamiltonian analytically averaged over the fast motions. This semi-numerical approach, inspired by \cite{Weinberg2007BarHaloInteraction}, allows us to understand the mechanism of dynamical friction in the general fast-slow regime from the fundamental viewpoint of nonlinear dynamics in a fully time-dependent manner. We then derive a complete analytic expression of dynamical feedback that incorporates contributions from both trapped and untrapped orbits. The latter is shown to be equivalent to the nonlinear-torque formula derived by \citetalias{Tremaine1984Dynamical}. We discuss its application to the Milky Way's bar.

This paper is structured as follows. In Section \ref{sec:model}, we introduce our bar-halo model.
In Section \ref{sec:phase_space_dynamics}, we describe the fundamental phase-space dynamics around a moving resonance using an angle-averaged Hamiltonian with an action variable comoving with the resonance. In Section \ref{sec:distribution_function}, we compute the time evolution of the distribution function in the slow angle-action plane and discuss the mechanism of dynamical friction and feedback in the general fast-slow regime. In Section \ref{sec:density_response}, we calculate the density response in real space and connect the phase-space dynamics to the generation of density wakes. In Section \ref{sec:dynamical_feedback}, we formulate dynamical feedback from both trapped and untrapped orbits using the angle-averaged distribution function $\fmix$. We then model the evolution of $\fmix$ using an extension of adiabatic theory and quantify dynamical feedback on the Galactic bar. We summarize our results in Section \ref{sec:conclusions}.

\section{Model}
\label{sec:model}

We restrict our analysis to the behaviour of test particles moving under the combined potential of the bar and the dark halo, i.e., we ignore the self-gravity of the halo's response. Following our previous work \citep{Chiba2022Oscillating}, we adopt a spherical isotropic halo with a \cite{Hernquist1990AnalyticalModel} profile:
\begin{align}
  \Phi_0(r) = - \dfrac{GM}{\rs+r}, ~~~~~ \rho_0(r) = \dfrac{M}{2\pi} \dfrac{\rs}{r(\rs+r)^3},  
  \label{eq:Hernquist}
\end{align}
where $G$ is the gravitational constant, $M = 1.5 \times 10^{12} \Msun$ is the total halo mass, and $\rs = 20 \kpc$ is the halo scale radius.

We perturb this halo with a quadrupole bar rotating at a time-dependent pattern speed $\Omegap(t)$:
\begin{align}
  \Phi_1(r,\vartheta,\varphi,t) = \Phib(r,t) \sin^2 \vartheta \cos 2 \left[\varphi - \int_0^t \drm t' \Omegap(t') \right],
  \label{eq:bar_potential}
\end{align}
where $(r,\vartheta,\varphi)$ are the spherical coordinates. The polar angle $\vartheta$ is measured from the pole, so the Galactic mid-plane is at $\vartheta = \pi/2$. Since the bar typically grows as it slows \citep[e.g.][]{debattista2000constraints,Martinez2006Evolution,VillaVargas2009Dark,Athanassoula2014Barslowdown,aumer2015origin,Fujii2019DryGalaxy}, we model the radial profile of the bar's potential such that the ratio between the bar's characteristic length $\rb$ and its corotation radius $\rCR$ is kept constant, that is, $b \equiv \rb/\rCR = {\rm const}$. This results in the following form \citep{Chiba2020ResonanceSweeping}:
\begin{align}
  \Phib(r,t) = - \dfrac{A \vc^2}{2} \left[\dfrac{r}{\rCR(t)}\right]^2 \left[\dfrac{b + 1}{b + r/\rCR(t)}\right]^5,
  \label{eq:bar_potential_amp}
\end{align}
where $A$ is the dimensionless strength of the bar and $\vc = 235 \kms$ is the local circular velocity. We adopt $A=0.02, b=0.28$ in approximation to the hydrodynamical model of \cite{Sormani2015GasIII} which is constrained by the kinematics of the gas in the inner Galactic disk.

Next we model the time-dependence of the bar's pattern speed $\Omegap(t)$. The simplest model one could think of is $\dOmegap = {\rm constant}$. However, $N$-body simulations typically find that the bar's slowing rate $-\dOmegap$ declines with decreasing $\Omegap$. Thus a better model would be to assume the following dimensionless quantity constant:
\begin{align}
  \eta \equiv - \dfrac{\dOmegap(t)}{\Omegap^2(t)} = {\rm const}.
  \label{eq:eta}
\end{align}
Integrating the above differential equation, we have
\begin{align}
  \Omegap(t) = \dfrac{\Omegapo}{1 + \eta \Omegapo t},
  \label{eq:bar_pattern_speed}
\end{align}
where $\Omegapo$ is the initial pattern speed. A constant $\eta$ thus implies a pattern speed evolving in inverse proportion to time\footnote{An alternative model would be to assume $\dOmegap/\Omegap$ constant, in which case the pattern speed declines exponentially.}.

\begin{figure}
  \begin{center}
    \includegraphics[width=8.5cm]{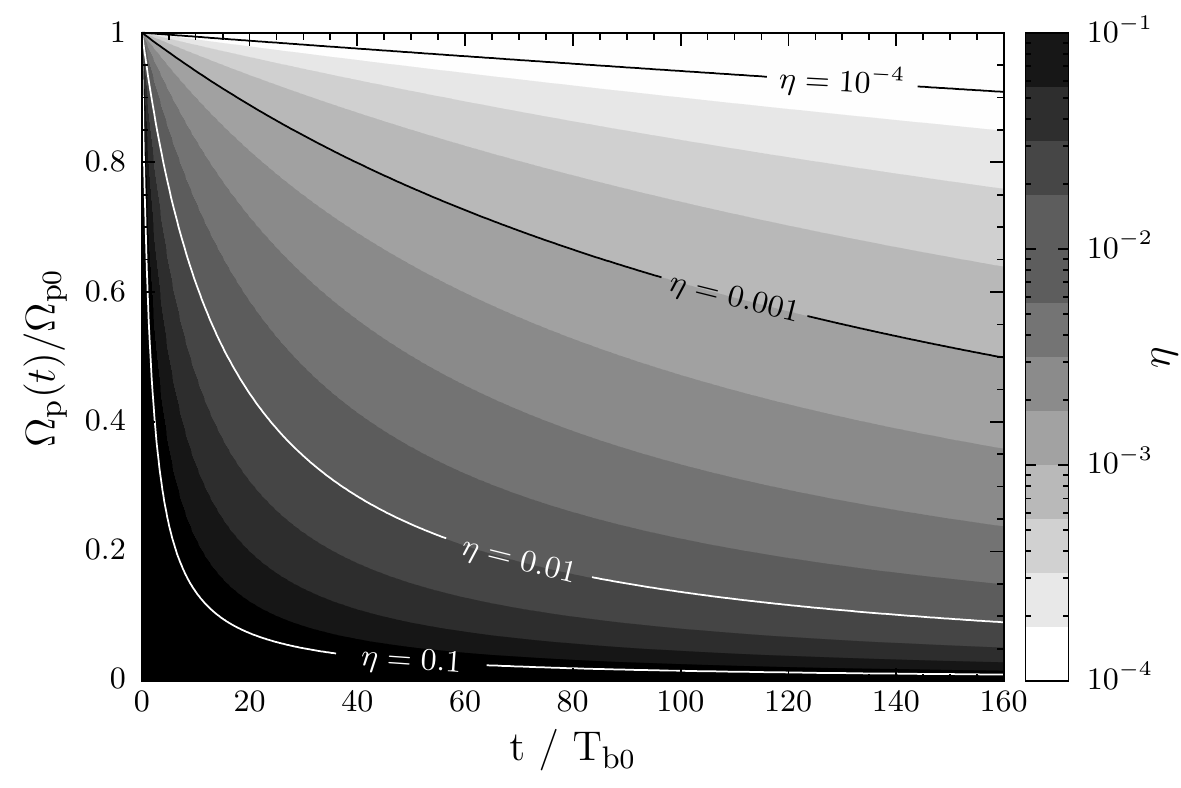}
    \caption{Model of the bar's pattern speed $\Omegap$ (equation \ref{eq:bar_pattern_speed}) as a function of time for various dimensionless slowing rates $\eta$. The axes are given in units of the bar's initial pattern speed $\Omegapo$ and rotation period $T_{\rm b0}=2\pi/\Omegapo$.}
    \label{fig:wp_eta}
  \end{center}
\end{figure}

Figure \ref{fig:wp_eta} shows the pattern speed curves for a range of slowing rate $\eta$. The time is given in units of the bar's initial rotation period $T_{\rm b0} = 2\pi/\Omegapo$. For the Milky Way's bar, its current pattern speed is measured to be around $\Omegapi = 33-41 \Gyr^{-1}$ \citep[][]{Portail2017Dynamical,Monari2019signatures,Sanders2019pattern,bovy2019life,Asano2020Trimodal,Binney2020Trapped,Chiba2021TreeRing,Kawata2021GalacticBarHotStar,Clarke2021ViracGaia,Li2022GasDynamics,Leung2022direct,Lucey2022Constraining,GaiaDR32022Mapping} and thus $T_{\rm b1} \sim 170 \pm 20 \Myr$. Suppose that the bar was born at a pattern speed twice as fast as its current value, so $T_{\rm b0} = T_{\rm b1}/2 = 85 \Myr$. Figure~\ref{fig:wp_eta} then shows that if $\eta = 0.001$ the bar would have slowed to its present state over $t \sim 160 T_{\rm b0} \sim 13.6 \Gyr$, while if $\eta = 0.01$ it merely took $t \sim 16 T_{\rm b0} \sim 1.36 \Gyr$. Thus, considering the age of the Galactic bar, a reasonable range for the slowing rate would be $\eta \in [0.001,0.01]$. Adopting a larger initial pattern speed will not change this estimate significantly, while a smaller initial pattern speed in principle allows for an arbitrary small $\eta$. Modelling of the local stellar kinematics suggests $\eta = 0.0036 \pm 0.0011$ \citep{Chiba2020ResonanceSweeping}.

\begin{figure}
  \begin{center}
    \includegraphics[width=8.5cm]{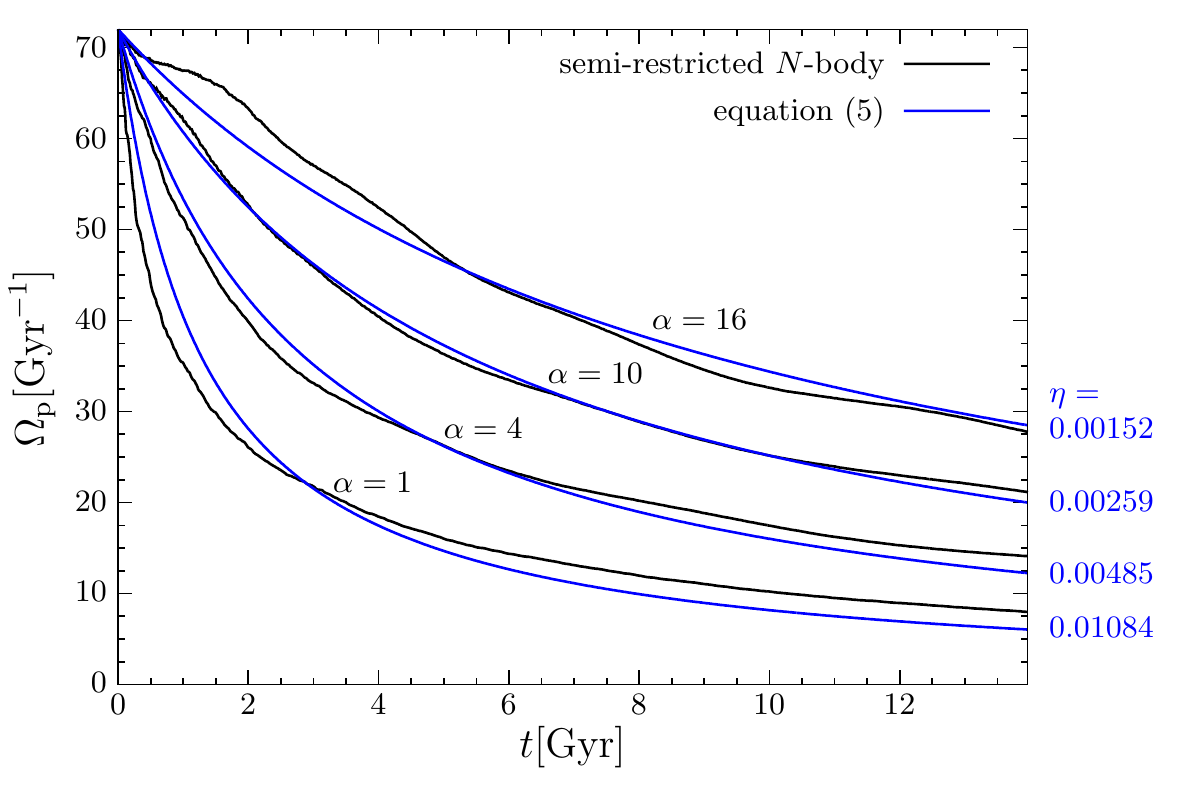}
    \caption{Evolution of the bar's pattern speed in semi-restricted $N$-body simulations (black) fitted with equation (\ref{eq:bar_pattern_speed}) (blue).
    }
    \label{fig:wp_t_n0}
  \end{center}
\end{figure}

To examine the adequacy of our model (\ref{eq:bar_pattern_speed}), and to clarify the conditions under which fast and slow evolutions are expected, we calculate the pattern speed by evolving the bar in accordance with the angular momentum transferred to the halo. The method is similar to the semi-restricted $N$-body simulation of \cite{Lin1983Numerical,weinberg1985evolution,Sellwood2006BarHaloI}, but we make the bar grow, which requires a time-dependent model for the bar's moment of inertia $I$. Following \cite{Beane2022StellarBars}, we let $I$ vary with the bar's corotation radius $\rCR$, but we choose the scaling $I \propto \rCR^3$ instead of $\rCR^2$ as the bar grows in mass as well as in length. Thus
\begin{align}
  \frac{I(t)}{\Irig} = \alpha \left[\frac{\rCR(t)}{r_{\rm CR1}}\right]^3,
  \label{eq:I}
\end{align}
where $\alpha$ is a free parameter and $r_{\rm CR1}=6.6 \kpc$ is the current corotation radius \citep{Chiba2021TreeRing,Clarke2021ViracGaia}. We have written $I$ in units of the $z$-moment of inertia of a rigid homogeneous ellipsoid \citep[e.g.][]{weinberg1985evolution} 
\begin{align}
  \Irig = \frac{\Mb}{5}(a_1^2+a_2^2),
  \label{eq:I_rigid}
\end{align}
with current bar properties: bar semi-major axis $a_1=5 \kpc$, minor axis $a_2=2 \kpc$, and bar mass $\Mb=10^{10} \Msun$ \citep{wegg2015structure}, resulting in $\Irig \simeq 6 \times 10^{10} \Msun \kpc^2$. The parameter $\alpha$ thus roughly measures the degree of deviation from a rigid bar. We assume that the change in the bar's moment of inertia is due to the gain of mass from the stellar disk, not due to dynamical friction. Hence, the term $\dot{I} \Omegap$ is excluded from the bar's equation of motion, $I \dOmegap = \tau$, where $\tau$ is dynamical friction. While dynamical friction may deform the bar and change $I$, we ignore this effect here.

The black curves in Fig.~\ref{fig:wp_t_n0} show the evolution of the bar pattern speed with $\alpha=1,4,10,16$. We choose an initial pattern speed $\Omega_{\rm p0} = 72 \Gyr^{-1}$. For all $\alpha$, our pattern speed model (blue) provides a reasonable fit to the simulations, where the best-fit $\eta$ is reported on the right. In agreement with \cite{weinberg1985evolution}, we find that a rigid bar ($\alpha=1$) slows significantly in a few bar rotation periods. \cite{Beane2022StellarBars} report $\alpha=4$ from their $N$-body+hydrodynamical simulation. The evolution with $\alpha=4$ is still fast, which is in line with $N$-body simulations without gas. A number of theories have been proposed to prevent such a rapid slowdown: (i) the dark halo has a low central density \citep{debattista2000constraints} or is non-existent \citep{Ghafourian2020Modified}, (ii) the halo has a net spin \citep{Long2014Secular,Fujii2019DryGalaxy}, or (iii) the gas provides angular momentum to the bar \citep{Bournaud2005lifetime,VillaVargas2010Gas,Hopkins2011analytic,Athanassoula2013BarGas}. The last point was recently revisited by \cite{Beane2022StellarBars} who showed that bars can spin down at an arbitrary low rate within a realistic fraction of gas in the disk ($4-16\%$). We therefore explore dynamical friction over a range of slowing rate, $\eta \in [0.001,0.032]$, where it is to be understood that a fast bar evolution $(\eta \sim 0.01)$ is expected in gas-poor galaxies, while a slow bar evolution $(\eta \sim 0.001)$ requires a certain supply of angular momentum from the gas.

\section{Phase-space dynamics around a moving resonance}
\label{sec:phase_space_dynamics}

In this section, we analyse the phase-space dynamics in the neighbourhood of a resonance moving at an arbitrary speed. The derivation up to equation (\ref{eq:H_averaged_slow}) can be found elsewhere \citep[e.g.][]{Tremaine1984Dynamical,binney2017orbital}, although we repeat them here for completeness.

To simplify the dynamics, we introduce angle-action variables $(\vtheta,\vJ)$, which are the natural coordinates to describe near-integrable systems \citep[e.g.][]{binney2008galactic}. By construction, the Hamiltonian of an integrable orbit is a function of the actions alone $H_0=H_0(\vJ)$, and thus the equations of motion take the simple form: $\dot{\vJ} = -\pd H_0/\pd \vtheta = {\bm 0}$ and $\dot{\vtheta} = \pd H_0/\pd \vJ \equiv \vOmega(\vJ) = {\rm constant}$. Actions are not only integrals of motion, but have the important property of being adiabatic invariants, making them especially valuable in describing the long-term evolution of galaxies.

In our model, the spherical symmetry and time symmetry of the unperturbed potential (equation \ref{eq:Hernquist}) guarantee the existence of a single set of angle-action coordinates that covers the entire phase space\footnote{The Hamiltonian of a general axisymmetric potential is not guaranteed to possess a global system of angle-action coordinates. However, an integrable Hamiltonian that closely approximates the true Hamiltonian can be constructed with the torus mapping technique introduced by \cite{McGill1990Torus}. \cite{binney2017orbital} applied this technique to model the bar's perturbation on the stellar disk in a realistic three-dimensional galactic potential.}. There are some freedom in the choice of the actions; we choose $\vJ=(\Jr,L,\Lz)$ where $\Jr$ is the radial action, $L$ is the norm of the angular momentum vector, and $\Lz$ is its $z$ component. The conjugate angles $\vtheta=(\thetar,\thetapsi,\thetaphi)$ describe the phase of radial oscillation, the phase of azimuthal motion in the orbital plane, and the longitude of the ascending node, which is a constant of motion.

Since the angle variables are $2\pi$ periodic, we can expand the perturbed potential (equation \ref{eq:bar_potential}) into a Fourier series
\begin{align}
  H(\vtheta, \vJ, t)
  &= H_0(\vJ) + \sum_{\bm n} \Phi_{\bm n}(\vJ,t) ~\e^{i \left[{\bm n} \cdot \vtheta - n_\varphi \int \!\drm t' \Omegap(t')\right]},
  \label{eq:H_FourierExpand}
\end{align}
where ${\bm n} \equiv (n_r,n_\psi,n_\varphi)$ is a triple of integers, and we have extracted the rotational time-dependence of the perturbation from the coefficients $\Phi_{\bm n}$ and coupled it to the exponential. $\Phi_{\bm n}$ is given in equation (B3) of \cite{Chiba2022Oscillating}. Near a resonance
\begin{align}
  \vN \cdot \vOmega(\vJ) - \Nphi \Omegap(t) = 0,
  \label{eq:resonance}
\end{align}
where $\vN \equiv (\Nr,\Npsi,\Nphi)$\footnote{Without loss of generality, we assume $\Nphi > 0$.}, the equations of motion generated by $H$ involve a term with ${\bm n} = \vN$ that evolves slowly, while others (${\bm n} \neq \vN$) rapidly alternate. This timescale separation suggests that we may capture the secular dynamics near resonances by averaging $H$ over the fast motions and retaining only the resonant term. Following the usual procedure, we exploit this timescale disparity by making a canonical transformation to the fast-slow angle-action variables $(\vtheta',\vJ')=(\thetafo,\thetaft,\thetas,\Jfo,\Jft,\Js)$, specific to each resonance $\vN$ \citepalias{Tremaine1984Dynamical},
\begin{align}
  &\thetafo = \thetar, \hspace{7.mm} \thetaft = \thetapsi, \hspace{7mm} \thetas = \vN \cdot \vtheta - \Nphi \int \!\drm t ~ \Omegap(t),
  \label{eq:slowfast_angle} \\
  &\Jfo = \Jr - \frac{\Nr}{\Nphi}\Lz, \hspace{5mm} \Jft = L - \frac{\Npsi}{\Nphi}\Lz, \hspace{5mm} \Js  = \frac{\Lz}{\Nphi},
  \label{eq:slowfast_action}
\end{align}
using the generating function of the second kind
\begin{align}
  S(\vtheta,\vJ',t) = \thetar \Jfo + \thetapsi \Jft + \left[\vN \cdot \vtheta - \Nphi \int \!\drm t ~ \Omegap(t)\right] \Js.
\label{eq:generating_function}
\end{align}
We write the new Hamiltonian as
\begin{align}
  H'(\vtheta',\vJ',t)
  &= H'_0(\vJ',t) + \sum_{\vk} \Psik(\vJ',t) ~\e^{i \vk \cdot \vtheta'},
  \label{eq:H_fastslow}
\end{align}
where the perturbation is expanded in the new angles $\vtheta'$ with new indices $\vk \equiv (\kfo,\kft,\ks)$. The unperturbed Hamiltonian acquires an explicit time dependence due to the varying pattern speed
\begin{align}
  H'_0(\vJ',t) = H_0(\vJ') + \frac{\pd S}{\pd t} = H_0(\vJ') - \Nphi \Omegap(t) \Js.
  \label{eq:H_fastslow_unperturbed}
\end{align}
Averaging equation (\ref{eq:H_fastslow}) over the fast angles $\vthetaf \equiv (\thetafo,\thetaft)$ yields
\begin{align}
  \bH(\thetas,\vJ',t)
  &= \frac{1}{(2\pi)^2} \int \drm^2 \vthetaf H'(\vtheta',\vJ',t) \nonumber \\
  &= \bH_0(\vJ',t) + \sum_{\ks} \Psi_{\ks}(\vJ',t) ~\e^{i \ks \thetas},
  \label{eq:H_averaged}
\end{align}
where $\bH_0 = H'_0$ and $\Psi_{\ks} \equiv \Psi_{(0,0,\ks)}$. For the dominant resonances with $\Nphi=2$, it can be shown that terms other than $\ks = \pm 1$ vanish \citep[][]{Chiba2022Oscillating}. Hence, we have
\begin{align}
  \bH(\thetas,\Js,t) = \bH_0(\Js,t) + \Psi(\Js,t) \cos \left(\thetas - \thetasres\right),
  \label{eq:H_averaged_slow}
\end{align}
where $\Psi \equiv 2 |\Psi_{(0,0,1)}|$ \citep[][Appendix B]{Chiba2022Oscillating}. For brevity, we have omitted the dependence on the constant fast actions $\vJf \equiv (\Jfo,\Jft)$, which may be regarded as parameters rather than dynamical variables.

Our focus here is on the impact of the time dependence of $\Omegap(t)$ and the associated growth of the bar $\Psi(\Js,t)$. The evolution of resonant systems in the classical adiabatic limit is described in detail by \cite{Neishtadt1975Passage} and \cite{Henrard1982Capture}. In the following, we summarize their key results, which we further develop in the subsequent Section \ref{sec:adiabatic_theory_comoving}.

\begin{figure*}
  \begin{center}
    \includegraphics[width=17.5cm]{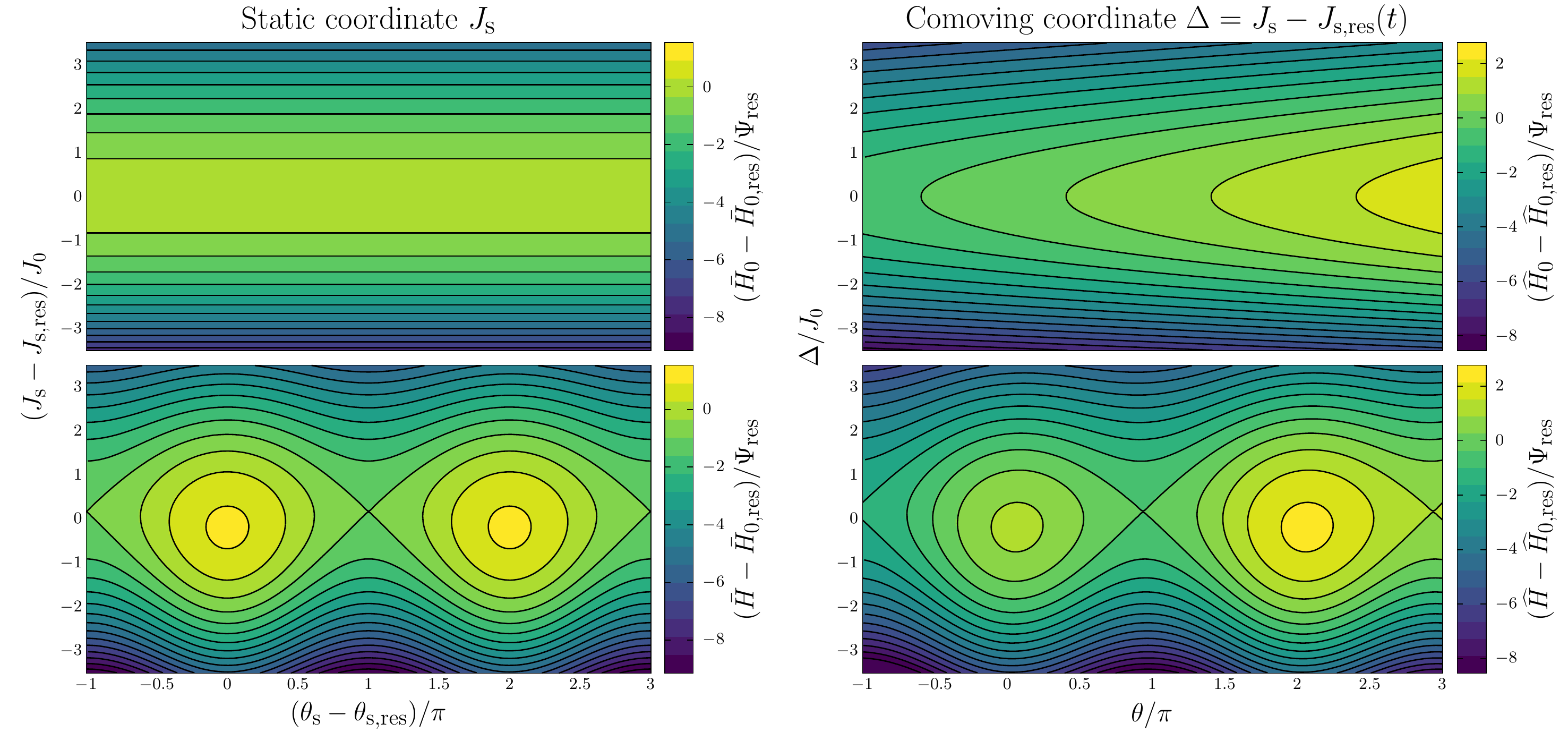}
    \caption{Level curves of the averaged Hamiltonian (bottom) and its unperturbed component (top). Left column: Hamiltonian $\bH$ (equation \ref{eq:H_averaged_slow}) in the static action coordinate $\Js$. Right column: Hamiltonian $\hH$ (equation \ref{eq:H_averaged_comoving}) in the comoving action coordinate $\Delta=\Js-\Jsres(t)$ when $\eta = 0.002$. $J_0$ is the characteristic width of the resonance defined in equation (\ref{eq:w0_J0}).}
    \label{fig:H_thetaJ}
  \end{center}
\end{figure*}

\subsection{Adiabatic evolution near a resonance}
\label{sec:adiabatic_theory}

The slow evolution of mechanical systems is best described in terms of adiabatic invariants. In the classical studies of resonant systems, adiabatic invariants are found by observing the phase-space trajectories in the `time-frozen' Hamiltonian of form (\ref{eq:H_averaged_slow}). Figure \ref{fig:H_thetaJ} left column shows the level curves of $\bH$ in the slow plane $(\thetas,\Js)$ at some instance of time. We chose the corotation resonance $\vN = (0,2,2)$ at $\vJf = {\bm 0}$\footnote{For the corotation resonance, $\vJf={\bm 0}$ corresponds to unperturbed orbits that are circular ($\Jr=0$) with zero inclination ($\Lz = L$).} as an example, but the discussion is general. The top-left figure shows the unperturbed Hamiltonian $\bH_0$. The trajectories are everywhere straight, and the Hamiltonian peaks exactly at the resonance $\Js=\Jsres$ because the slope $\pd \bH_0 / \pd \Js = \vN \cdot \vOmega - \Nphi \Omegap$ vanishes there. The bottom-left figure shows the contours of the perturbed Hamiltonian $\bH$. Because $\bH_0$ takes an extremum at $\Jsres$, the presence of even the smallest perturbation can change the topology of the phase-space trajectories near the resonance: the level curves of $\bH$ connect phase space above and below the resonance, resulting in trajectories that oscillate in $\thetas$. This is the regime of \textit{libration}. Librating orbits are referred to as being trapped in resonance. Untrapped orbits, in contrast, remain on either side of the resonance and freely \textit{circulate}, i.e., their slow angle evolves monotonically (albeit not linearly in time).

The adiabatic invariant in the perturbed system is no longer $\Js$ but the phase-space area enclosed by the contours of the time-frozen $\bH$ 
\begin{align}
  \Jl = \frac{1}{2 \pi} \oint \drm \thetas \Js,
  \label{eq:new_action}
\end{align}
where we have divided the area by $2 \pi$ to make it an action variable. The angles $\thetal$ conjugate to $\Jl$ describe the phase of libration (or circulation). Adiabatic theory assumes that $\bH$ evolves slowly enough that actions $\Jl$ of most orbits are approximately preserved. The adiabaticity, however, inevitably breaks down at the separatrix where the orbital period becomes infinite. There, orbits may transit from one regime of phase space to another and the actions $\Jl$ change discontinuously. The fate of an orbit crossing the separatrix is known to be extremely sensitive to its angle $\thetal$ and thus cannot be practically predicted. One can, however, calculate the probability of transition averaged over $\thetal$ \citep{Neishtadt1975Passage,Henrard1982Capture}, and this has been used to model the angle-averaged (coarse-grained) distribution function of stellar systems near resonances \citep{Sridhar1996Adiabatic}. We will come back to this point in Section \ref{sec:dynamical_feedback} when we model dynamical feedback.

The adiabatic assumption requires the typical libration period $\Tl$ to be sufficiently shorter than the timescale at which the separatrices move by the width of the trapped phase-space. Let $v_{\pm}$ be the speed of the separatrices above and below the resonance. To order of magnitude, $v_{\pm}$ can be written as the sum of the rate at which the resonance moves $\dJsres$ and the rate at which the width of the trapped region $\dJres$ changes:
\begin{align}
  v_{\pm} \sim \dJsres \pm \ddJres = \left(1 \pm \gamma \right) \dJsres,
  \label{eq:speed_of_separatrix}
\end{align}
where the dimensionless parameter
\begin{align}
  \gamma \equiv \frac{\ddJres}{\dJsres}
  \label{eq:gamma}
\end{align}
describes the change in resonant width with respect to the change in resonant position. The adiabatic condition requires
\begin{align}
  \Tl \ll \frac{\delta J_{\rm res}}{{\rm max}(v_+,v_-)} = \frac{\delta J_{\rm res}}{\left(1 + |\gamma|\right) |\dJsres|}.
  \label{eq:adiabatic_condition}
\end{align}
As we shall see later (equations \ref{eq:w0_J0}, \ref{eq:dJsresdt}), the libration period generally scales as $\Tl \sim 1/\sqrt{A}$, where $A$ is the order of perturbation (equation \ref{eq:bar_potential_amp}), the resonance width scales as $\delta J_{\rm res} \sim \sqrt{A}$, while the speed of the resonance as $|\dJsres| \sim \eta$ \citep[see also][Section 2.4]{Chiba2020ResonanceSweeping}. Therefore, to leading order, the adiabatic assumption requires
\begin{align}
  \left(1+|\gamma|\right) \frac{\eta}{A} \ll 1.
  \label{eq:adiabatic_condition_simple}
\end{align}
For the problem at hand, $\gamma$ is typically of order $10^{-1}$ over the majority of phase space (Appendix \ref{sec:G_Gamma}), so the adiabatic condition is largely dictated by the ratio $\eta/A$. Models of the Galactic bar suggest $A \sim 10^{-2}$ while $\eta \sim 10^{-3}$ (Section \ref{sec:model}), so $\eta / A \sim 10^{-1}$. Hence, in real bar-halo systems, the adiabatic criterion (\ref{eq:adiabatic_condition_simple}) is unlikely to be sufficiently satisfied.

\subsection{Adiabatic theory in the comoving frame}
\label{sec:adiabatic_theory_comoving}

In this section, we show that the range of applicability of the adiabatic theory can be extended -- the adiabatic condition can be eased -- by making a simple canonical transformation.

As discussed in the previous Section, the adiabatic criterion depends on the speed with which the separatrix moves and is determined predominantly by the change in the position of the resonance rather than the accompanying change in the shape of the separatrix. This suggests that we may find a phase-space coordinate in which the separatrix appears more stationary by making a canonical transformation to a slow action variable comoving with the sweeping resonance
\begin{align}
  \Delta \equiv \Js - \Jsres(t).
  \label{eq:Delta_Js}
\end{align}
The slow angle need not be changed, but for brevity, we redefine
\begin{align}
  \theta \equiv \thetas - \thetasres.
  \label{eq:Delta_thetas}
\end{align}
A suitable generating function for this transformation is
\begin{align}
  S(\thetas, \Delta, t) = \left(\thetas - \thetasres\right) \left[\Delta + \Jsres(t)\right].
  \label{eq:generating_func}
\end{align}
The new Hamiltonian is then
\begin{align}
  \hH(\theta,\Delta,t) = \hH_0(\theta,\Delta,t) + \Psi(\Delta,t) \cos \theta,
  \label{eq:H_averaged_comoving}
\end{align}
where the unperturbed part now has an additional term that is linear in the slow angle $\theta$:
\begin{align}
  \hH_0(\theta,\Delta,t) = \bH_0 + \frac{\pd S}{\pd t} = \bH_0[\Delta+\Jsres(t),t] + \dJsres(t) \theta.
  \label{eq:H_averaged_comoving_unperturbed}
\end{align}
This new unperturbed Hamiltonian $\hH_0$ has lost periodicity in the slow angle because it returns the value of the old Hamiltonian $\bH_0$ along lines of constant comoving action $\Delta$.

The right column of Fig.~\ref{fig:H_thetaJ} shows the change in the level curves of the time-frozen Hamiltonian caused by this canonical transformation. Note that the axes of the left and right columns remain the same, although different labels are used to highlight the change in canonical coordinates. In the new comoving coordinates, the unperturbed trajectories (top-right figure) no longer appear straight but move towards low $\Delta$ because the resonance is moving up. The sign of $\dot{\theta}$ switches from negative to positive as the orbits pass the resonance. As with the case in the static coordinate, the presence of the perturbation (bottom-right) changes the topology near the resonance, but crucially, the phase space of libration has diminished in size. In addition, the libration is no longer symmetric around $\theta=0$: the angles of the stable fixed points (the local extrema of $\hH$) have shifted positively, while that of the unstable fixed points (the saddle points) have shifted negatively. Unlike the trapped orbits, the phase curves of the untrapped orbits are open: they approach the resonance from above, make a U-turn at the resonance, and transition to the lower circulating regime.

\begin{figure}
  \begin{center}
    \includegraphics[width=8.5cm]{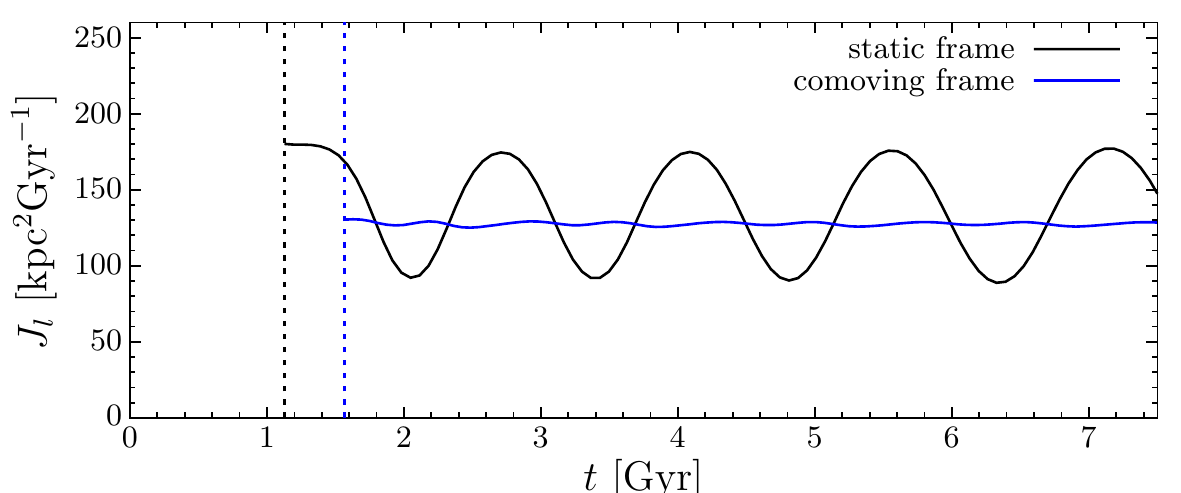}
    \caption{Time variation of the libration action of an orbit trapped and dragged by a moving resonance. The bar slowed from $\Omegap=60$ to $30 \Gyr^{-1}$ in approximately $7.5 \Gyr$ (slowing rate $\eta = 0.002$) and the orbit was initially placed at $\Js = 550 \kpc^2 \Gyr^{-1}$. The libration action calculated along the contours of the time-frozen Hamiltonian in the comoving frame $\hH$ (blue) is better preserved than that in the static frame $\bH$ (black).}
    \label{fig:Jl}
  \end{center}
\end{figure}

The phase-space area enclosed by the trapped orbits in the comoving frame defines a new action of libration 
\begin{align}
  \Jl \equiv \frac{1}{2\pi} \oint \drm \theta \Delta,
  \label{eq:Jl}
\end{align}
where the integral runs along constant time-frozen $\hH$. Figure~\ref{fig:Jl} compares the variation of the action of libration defined by the time-frozen $\bH$ (static action coordinate) with that defined by the time-frozen $\hH$ (comoving action coordinate). Specifically, we integrated a single trapped orbit in the time-dependent Hamiltonian (here, static or comoving makes no difference), and for each phase point $\bm{w}_i$ at each time step, we calculated the libration action by integrating a new orbit starting from $\bm{w}_i$ but in the time-frozen Hamiltonian (here, static or comoving is relevant). Clearly, the libration action in the comoving frame is better preserved. The dotted lines mark the time when the orbit enters the region of libration (defined by the time-frozen $\bH$ and $\hH$ respectively), which is slightly earlier in the static case because the trapped region is larger (cf. Fig.~\ref{fig:H_thetaJ}).

The reason for the improvement in the conservation of $\Jl$ can be understood by re-evaluating the adiabatic condition. In the comoving frame, the resonance is fixed, so the speed of the separatrix is given solely by the change in shape of the trapped region
\begin{align}
  v_{\pm} \sim \pm \ddJres = \pm \gamma \dJsres.
  \label{eq:speed_of_separatrix_comove}
\end{align}
The adiabatic condition is then 
\begin{align}
  \Tl \ll \frac{\delta J_{\rm res}}{{\rm max}(v_+,v_-)} = \frac{\delta J_{\rm res}}{|\gamma \dJsres|},
  \label{eq:adiabatic_condition_comove}
\end{align}
which, to leading order, requires 
\begin{align}
  |\gamma| \frac{\eta}{A} \ll 1.
  \label{eq:adiabatic_condition_comove_simple}
\end{align}
As compared to the original criterion (\ref{eq:adiabatic_condition_simple}), the new criterion depends directly on the size of the parameter $\gamma$ (equation \ref{eq:gamma}). This implies that if the shape of the separatrix in the comoving frame varies little ($\gamma \ll 1$), the adiabaticity does not necessary require a slowly moving resonance $(\eta/A \ll 1)$ as in the classical adiabatic assumption. In the example given above (Fig.~\ref{fig:Jl}), $\gamma \sim 0.1$ so $\gamma \eta / A \sim 0.01$, which satisfies the adiabatic criterion better than in the static coordinate by an order of magnitude. While the degree of improvement depends on the chosen orbit, the evolution is always more adiabatic in the comoving frame. This implies that the instantaneous phase flow is better represented by the level curves of $\hH$ rather than that of $\bH$. We will demonstrate in Section \ref{sec:distribution_function} that the phase-space flow identified in the comoving action coordinate indeed describes the numerical simulations remarkably well.

To gain further insights into the phase-space structure of $\hH$, let us employ the standard pendulum approximation \citep[e.g.][]{lichtenberg1992regular}. That is, we Taylor expand $\hH$ in $\Delta$ up to second order for the unperturbed term and to zeroth order for the perturbed term
\begin{align}
  \hH(\theta,\Delta,t) 
  &\simeq \left.\bH_0\right\vert_{\Delta=0} + \left.\frac{\pd \bH_0}{\pd \Delta}\right\vert_{\Delta=0} \Delta + \frac{1}{2} \left.\frac{\pd^2 \bH_0}{\pd \Delta^2}\right\vert_{\Delta=0} \Delta^2 \nonumber \\
  &+ \left.\Psi\right\vert_{\Delta=0} \cos \theta + \dJsres \theta.
  \label{eq:H_averaged_Taylor_expansion}
\end{align}
The first term can be removed as it does not affect the slow dynamics, and the second term is zero at the resonance. Hence
\begin{align}
  \hH(\theta,\Delta,t)
  &= \frac{1}{2} G \Delta^2 - F \cos \theta + \dJsres \theta,
  \label{eq:H_averaged_comoving_Taylor}
\end{align}
where
\begin{align}
  G \equiv \left.\frac{\pd^2 \bH_0}{\pd \Delta^2}\right\vert_{\Delta=0}=\left.\frac{\pd \left(\vN \cdot \vOmega \right)}{\pd \Js}\right\vert_{\Js=\Jsres} ~~~ {\rm and} ~~~ F \equiv -\left.\Psi\right\vert_{\Delta=0}.
  \label{eq:GF}
\end{align}
The Hamiltonian (\ref{eq:H_averaged_comoving_Taylor}) is equivalent to that of a pendulum subject to an external torque. For the main bar resonances, the quantity $G$ is negative in much of the phase space (Appendix \ref{sec:G_Gamma}). $F$ is defined to be negative such that the Hamiltonian resembles that of a real pendulum. In the following, we thus assume $FG>0$. Cases with $FG<0$ can be dealt similarly by shifting the angle as appropriate. The two parameters together define two important scales for the motion near a resonance:
\begin{align}
  \omega_0 = \sqrt{FG}, ~J_0 = \sqrt{F/G} ~~~\leftrightarrow~~~ F = - \omega_0 J_0, ~G = - \omega_0 / J_0.
  \label{eq:w0_J0}
\end{align}
When the resonance is static ($\dJsres = 0$), $\omega_0$ describes the libration frequency of orbits with infinitesimal libration amplitude and thus provides the timescale of libration, while $J_0$ is one-half of the maximum libration amplitude and thus characterizes the size of the trapped region \citep{lichtenberg1992regular}. Note that when $\dJsres \neq 0$, both $G$ and $F$, as well as the two scale parameters, are time dependent because they are evaluated on the moving resonance. 

Using equation (\ref{eq:w0_J0}), we may rewrite the Hamiltonian (\ref{eq:H_averaged_comoving_Taylor}) as 
\begin{align}
  \hH(\theta,\Delta,t) = \frac{1}{2} G \Delta^2 - F \left(\cos \theta + s \theta \right),
  \label{eq:H_target}
\end{align}
where 
\begin{align}
  s \equiv \frac{\dJsres}{\omega_0 J_0} = \frac{\Tl}{\Td}
  \label{eq:speed_parameter}
\end{align}
is a dimensionless parameter that describes the ratio of the characteristic libration time $\Tl=\omega_0^{-1}$ to the time the resonance takes to drift by its width $\Td = J_0 / \dJsres$. This parameter $s$ is in fact the dimensionless \textit{speed} parameter introduced in \citetalias{Tremaine1984Dynamical}. To demonstrate this, we derive $\dJsres$ by differentiating the resonance condition (\ref{eq:resonance}) with time
\begin{align}
  0 &= \frac{\drm}{\drm t} \left[ \vN \cdot \vOmega\left(\Jsres\right) - \Nphi \Omegap \right], \nonumber \\
  &= \dJsres \left.\frac{\pd \left(\vN \cdot \vOmega \right)}{\pd \Js}\right\vert_{\Js=\Jsres} - \Nphi \dOmegap.
  \label{eq:dJsresdt_derivation}
\end{align}
Using $G$ (equation \ref{eq:GF}), we obtain
\begin{align}
  \dJsres = \frac{\Nphi \dOmegap}{G}.
  \label{eq:dJsresdt}
\end{align}
Substituting this into equation (\ref{eq:speed_parameter}), we have 
\begin{align}
  s = - \frac{\Nphi \dOmegap}{GF},
  \label{eq:speed_parameter_2}
\end{align}
which is equivalent to equation (89) of \citetalias{Tremaine1984Dynamical}. This expression indicates that, as opposed to the dimensionless slowing rate $\eta$ (equation \ref{eq:eta}), which measures $\dOmegap$ with respect to the bar's dynamical time $\Omegap^{-1}$, the speed parameter $s$ measures $\dOmegap$ with respect to the characteristic libration time $\omega_0^{-1}=(GF)^{\-1/2}$. Hence, while $\eta$ is a global parameter, $s$ is a local parameter that depends on the resonance $\vN$ as well as on the position on each resonance set by the fast actions $\vJf$.

To clarify the factors featuring the Hamiltonian (\ref{eq:H_target}), we further rewrite it in a dimensionless form:
\begin{align}
  \hH = F \left[\frac{1}{2} \left(\frac{\Delta}{J_0}\right)^2 + V(\theta) \right],
  \label{eq:H_target_dimensionless}
\end{align}
where $V$ is the dimensionless effective potential
\begin{align}
  V(\theta) = - \cos \theta - s \theta.
  \label{eq:V}
\end{align}
The phase-space structure of $\hH$ is thus characterized solely by $s$.

\begin{figure}
  \begin{center}
    \includegraphics[width=8.5cm]{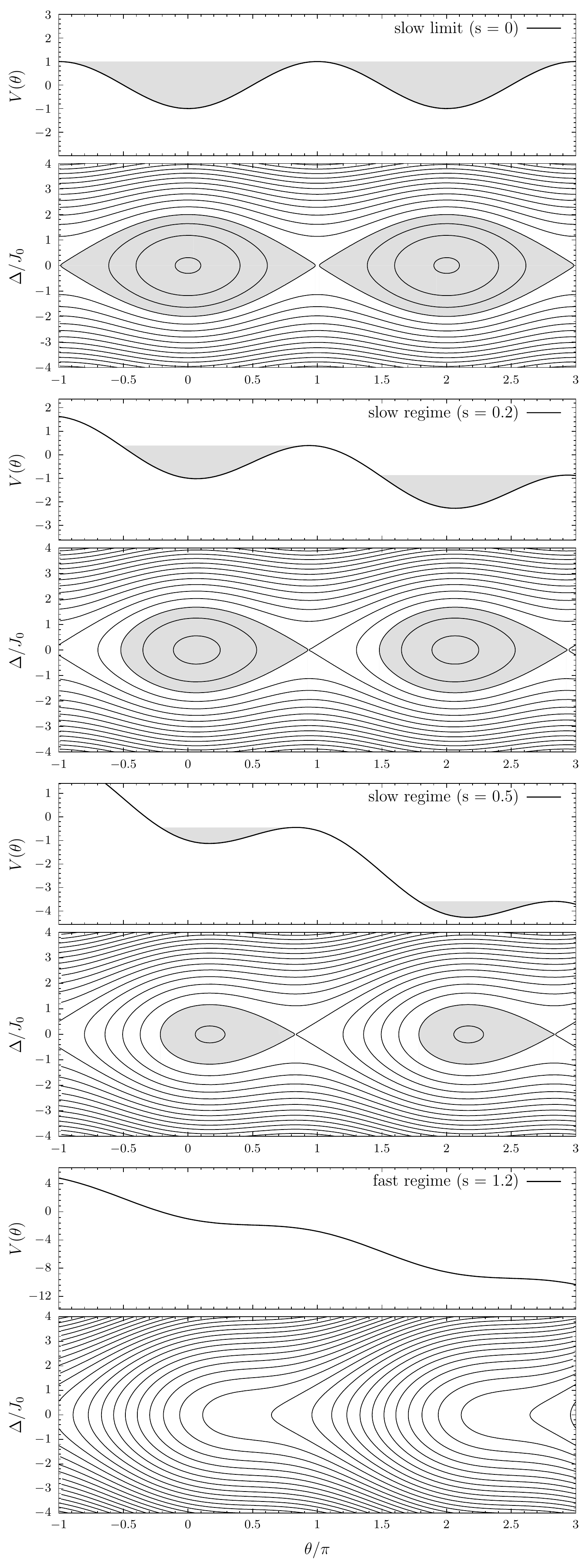}
    \caption{The effective potential of the resonance (\ref{eq:V}) and the contours of the Hamiltonian (\ref{eq:H_target_dimensionless}) for various speed $s$ increasing from top to bottom.}
    \label{fig:H_ModPend}
  \end{center}
\end{figure}

Figure~\ref{fig:H_ModPend} shows the potential and the contours of the (time-frozen) Hamiltonian (\ref{eq:H_target_dimensionless}) in the slow plane for various $s$. A similar phase portrait has been observed in a variety of physical phenomena that involve a sweeping resonance: the radial migration of planet in proto-planetary disk \citep{Ogilvie2006PlanetaryMigration}, the propagation of waves in an inhomogeneous plasma \citep[e.g.][]{Shklyar2009Oblique,Artemyev2018Trapping}, and the acceleration of particles/molecules by chirped lasers \citep[e.g.][]{Kalyakin2008Asymptotic,Armon2016Capture}.

The top panel of Fig.~\ref{fig:H_ModPend} depicts the phase flow in the \textit{slow limit} ($s = 0$, constant pattern speed) which resembles that of a standard pendulum. The phase flow streams anticlockwise around the fixed points since $F < 0$. Unlike the original Hamiltonian (Fig.~\ref{fig:H_thetaJ}), this Hamiltonian is symmetric about the resonance $\Delta=0$ since we have neglected the high-order Taylor series.

\begin{figure}
  \begin{center}
    \includegraphics[width=8.5cm]{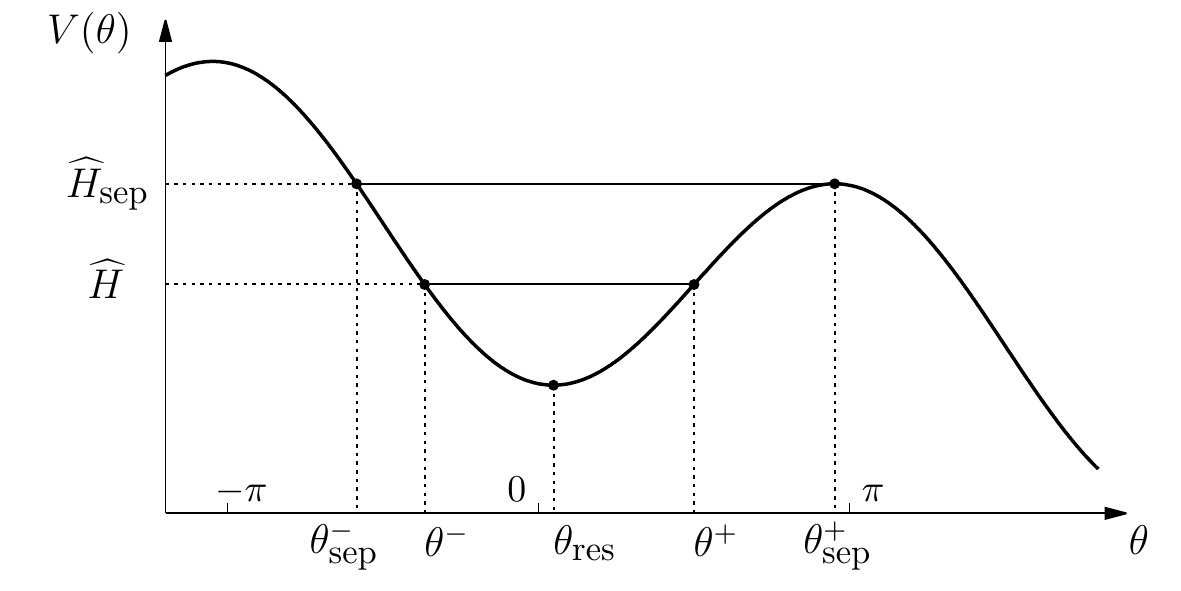}
    \caption{Schematic diagram of the effective potential of the resonance.}
    \label{fig:schematic_effective_potential}
  \end{center}
\end{figure}

The middle two panels of Fig.~\ref{fig:H_ModPend} show the phase flow in the \textit{slow regime} ($0 < s < 1$, resonance moving slowly). The potential gets tilted and the trapped phase-space (grey area) accordingly shrinks, allowing untrapped orbits to transit across the resonance. Only the trajectories of the trapped orbits are closed. The action of libration (the enclosed phase-space area) can be calculated by solving equation (\ref{eq:H_target_dimensionless}) for $\Delta$ and plugging it into equation (\ref{eq:Jl})
\begin{align}
  \Jl &= \frac{1}{2\pi} \oint \drm \theta \sqrt{2J_0^2\left(\hH/F + \cos \theta + s \theta \right)} \nonumber \\
  &= \frac{\sqrt{2} J_0}{\pi} \int_{\theta^{-}}^{\theta^{+}} \drm \theta \sqrt{\cos \theta - \cos \theta^{+} + s \left(\theta - \theta^{+} \right)}
  \label{eq:Jl_pend}
\end{align}
where $\theta^{\pm}$ are the roots of the equation $\hH + F (\cos \theta + s \theta) = 0$. Figure~\ref{fig:schematic_effective_potential} illustrates the definition of $\theta^{\pm}$.

The bottom panel of Fig.~\ref{fig:H_ModPend} presents the \textit{fast regime} $(s > 1)$. The effective potential no longer forms a local minima since 
\begin{align}
  \frac{\drm V}{\drm \theta} = \sin \theta - s < 0 ~~\forall~~ \theta,
  \label{eq:dVdtheta}
\end{align}
meaning that no orbits can stay trapped -- the resonance sweeps past the orbits before they can complete a full cycle of libration. As described in \citetalias{Tremaine1984Dynamical} and \cite{Chiba2022Oscillating} (Appendix D), this is the only regime where linear perturbation theory works near resonances beyond the libration time\footnote{Note though that other sources of time-dependent evolution in galaxies may disrupt motions of libration and hence allow the system to remain in the linear regime.}.

\begin{figure}
  \begin{center}
    \includegraphics[width=8.5cm]{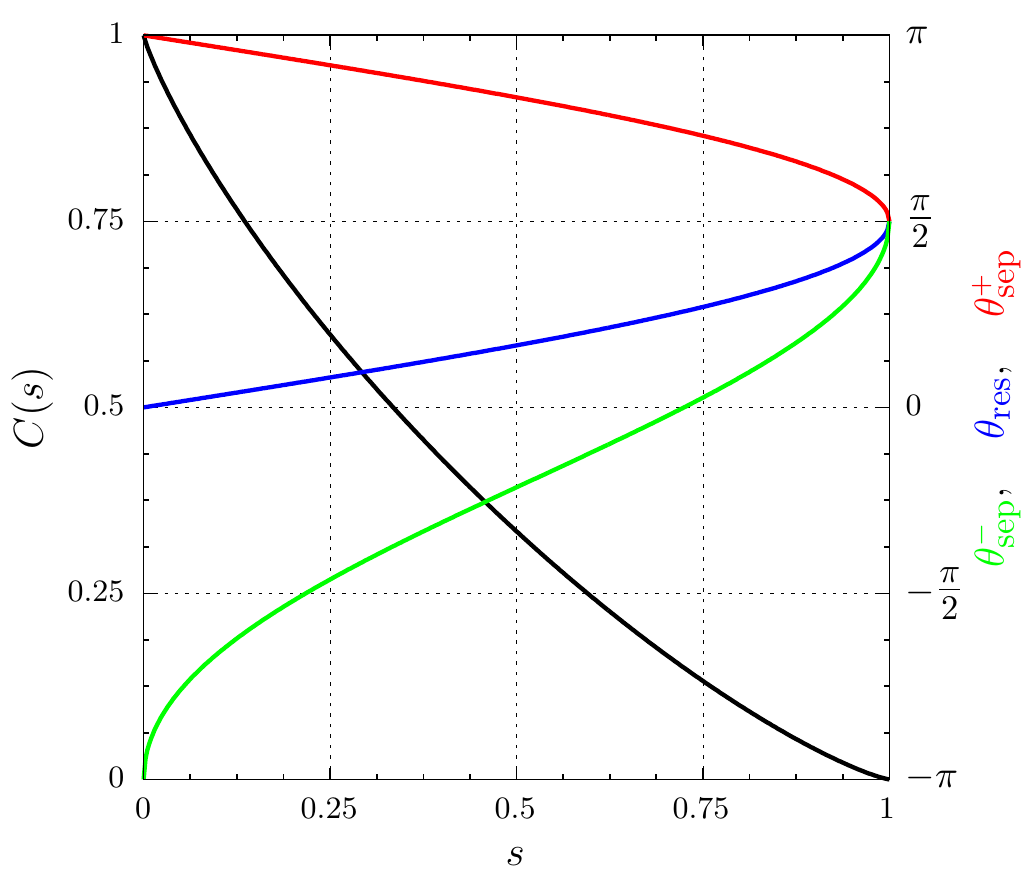}
    \caption{Black curve: The resonant-volume factor $C$ (equation \ref{eq:resonant_volume_factor}) as a function of the dimensionless speed $s$ (equation \ref{eq:speed_parameter}). Blue and red curves: The local minimum $\thetares$ and maximum $\thetasepp$ of the resonant potential $V$ in the range $0 < \thetares < \pi/2 < \thetasepp < \pi$. Green curve: The root $\thetasepm$ of the equation $V(\theta) = V(\thetasepp)$ closest to $\thetasepp$ (see Fig.~\ref{fig:schematic_effective_potential}). The difference $\Delta \theta = \thetasepp - \thetasepm$ is the angular width of the trapped phase-space.}
    \label{fig:resonant_volume_factor}
  \end{center}
\end{figure}

None of the trajectories presented in Fig.~\ref{fig:H_ModPend} escape or get captured into the resonance since we have frozen $\hH$ in time. For orbits to cross the separatrix, the volume of trapping must change. The trapped volume is given by the action of libration (equation \ref{eq:Jl_pend}) evaluated along the separatrix
\begin{align}
  \Jlsep = \frac{8 J_0}{\pi} C(s),
  \label{eq:Jlsep}
\end{align}
where the function $C(s)$ is the \textit{resonant-volume factor}\footnote{The function $C(s)$ is equivalent to 
\begin{align}
  Q(s) = \frac{\sqrt{2}}{\pi s} \int_{\theta^{-}_{\rm sep}}^{\theta^{+}_{\rm sep}} \drm \theta \sin \theta \sqrt{\cos \theta - \cos \theta^{+}_{\rm sep} + s \left(\theta - \theta^{+}_{\rm sep} \right)}
  \label{eq:Q}
\end{align}
derived in \citetalias[][]{Tremaine1984Dynamical} (equation 99) up to a constant factor because
\begin{align}
  4\sqrt{2}C(s) - \frac{\pi}{\sqrt{2}}Q(s)
  &= \frac{1}{s} \!\int_{\theta^{-}_{\rm sep}}^{\theta^{+}_{\rm sep}}\!\drm \theta \left(s \!-\! \sin \theta\right) \sqrt{\cos \theta \!-\! \cos \theta^{+}_{\rm sep} \!+\! s \!\left(\theta \!-\! \theta^{+}_{\rm sep} \right)} \nonumber \\
  &= \frac{1}{s}\left[\frac{2}{3}\left\{ \cos \theta - \cos \theta^{+}_{\rm sep} + s \left(\theta - \theta^{+}_{\rm sep} \right) \right\}^{\frac{3}{2}} \right]_{\theta^{-}_{\rm sep}}^{\theta^{+}_{\rm sep}} \nonumber \\
  &= 0, 
  \label{eq:C_Q}
\end{align}
where the last line holds since $V(\theta^{+}_{\rm sep}) = V(\theta^{-}_{\rm sep})$ (see Fig.~\ref{fig:schematic_effective_potential}).}
\begin{align}
  C(s) = \dfrac{1}{4\sqrt{2}} \int_{\theta^{-}_{\rm sep}}^{\theta^{+}_{\rm sep}} \drm \theta \sqrt{\cos \theta - \cos \theta^{+}_{\rm sep} + s \left(\theta - \theta^{+}_{\rm sep} \right)},
  \label{eq:resonant_volume_factor}
\end{align}
with domain $0 \leq s \leq 1$ and range $0 \leq C \leq 1$. The upper bound of the integral $\thetasepp$ is at the local maximum of the effective potential (Fig.~\ref{fig:schematic_effective_potential}), which corresponds to the saddle point of the Hamiltonian. Hence, $V'(\thetasepp) = \sin \theta^{+}_{\rm sep} - s = 0$, which gives
\begin{align}
  \thetasepp = \sin^{-1} s, ~~~~ \left( \frac{\pi}{2} \leq \thetasepp \leq \pi \right).
  \label{eq:thetasep}
\end{align}
The lower bound of the integral $\theta^{-}_{\rm sep}$ is given by the root of the equation $V(\theta) = V(\thetasepp)$ closest to $\thetasepp$. For subsequent use, we also define $\thetares$ as the angle at which the potential is locally minimum:
\begin{align}
  \thetares = \sin^{-1} s, ~~~~ \left( 0 \leq \thetares \leq \frac{\pi}{2} \right).
  \label{eq:thetares}
\end{align}

Figure~\ref{fig:resonant_volume_factor} plots the resonant-volume factor $C$ (black) as a function of the speed parameter $s$. The function is unity at $s=0$ and monotonically declines with increasing $s$ until it vanishes at $s=1$. We also plot the angles $\thetasepp$ (red) and $\thetasepm$ (green), which demarcate the trapped region. At $s=0$, these angles are separated by $2\pi$. As $s$ increases, they converge and eventually coincide at $s=1$, in line with the vanishing of $C$.

It is worth noting that as $s$ approaches unity, the bounding angles $\thetasep^{\pm}$ and the centre of libration $\thetares$ converge to $\pi/2$, not zero. This has an interesting observational implication: \textit{detecting a shift in the centre of libration allows us to directly measure $s$ and thus the bar's slowing rate $\dOmegap$}. For the corotation resonance, this amounts to measuring the angular deviation of the Lagrange points from the bar's axes, $\varphi_{\rm L} - \phib$, which yields the speed of the resonance as $s = \sin \left[2 \left( \varphi_{\rm L} - \phib \right) \right]$. We will pursue this possibility in a separate paper.

\section{Evolution of the distribution function}
\label{sec:distribution_function}

Using the averaged Hamiltonian derived in the previous section, we now study the evolution of the distribution function (DF) around a moving resonance. The modelled DF can then be used to derive integrated quantities such as the density response in real space (Section \ref{sec:density_response}).

The evolution of the DF $f(\vtheta,\vJ,t)$ is described by the collisionless Boltzmann equation (CBE) 
\begin{align}
  \frac{\drm f}{\drm t} = \frac{\pd f}{\pd t} + \frac{\pd f}{\pd \vtheta}\cdot\frac{\pd H}{\pd \vJ} - \frac{\pd f}{\pd \vJ}\cdot\frac{\pd H}{\pd \vtheta} = 0.
  \label{eq:CBE}
\end{align}
The CBE states that, unlike in a collisional model \citep[e.g.][]{Hamilton2022BarResonanceWithDiffusion}, the phase-space density along an orbit is conserved. Thus, we immediately have
\begin{align}
  f(\vtheta, \vJ, t) = f(\vtheta_0, \vJ_0, 0),
  \label{eq:CBE_sol}
\end{align}
and all we need is to calculate the phase-space trajectories under the modelled Hamiltonian. Since we do not model the system self-consistently, i.e., $H$ does not depend on $f$, determination of the trajectories can be done independently for each point in phase space.

In the averaged system $\bH$, the two fast actions $\vJf$ are conserved, so the trajectories only have four degrees of freedom. We further assume that the DF is initially fully phase-mixed in the angles. The DF then remains uniform in the fast angles $\vthetaf$ at all times, since the averaged trajectories are independent of $\vthetaf$. The problem then reduces to solving the trajectories in the 2D slow angle-action space
\begin{align}
  f(\thetas, \Js, t) = f(\theta_{\rm s 0}, J_{\rm s 0}, 0),
  \label{eq:CBE_sol_slow}
\end{align}
where, as with the Hamiltonian, we have omitted references to $\vJf$.

In analysing the dynamics, it is often sufficient to observe only a fraction of the entire phase space at a few snapshots. We take advantage of this by using the backward-integration technique \citep[e.g.][]{Vauterin1997construction,dehnen2000effect}. Integrating the trajectories backward in time saves significant numerical cost because we need only to integrate trajectories that pass the phase space of interest at the time of taking the snapshot.

\begin{figure}
  \begin{center}
      \includegraphics[width=8.5cm]{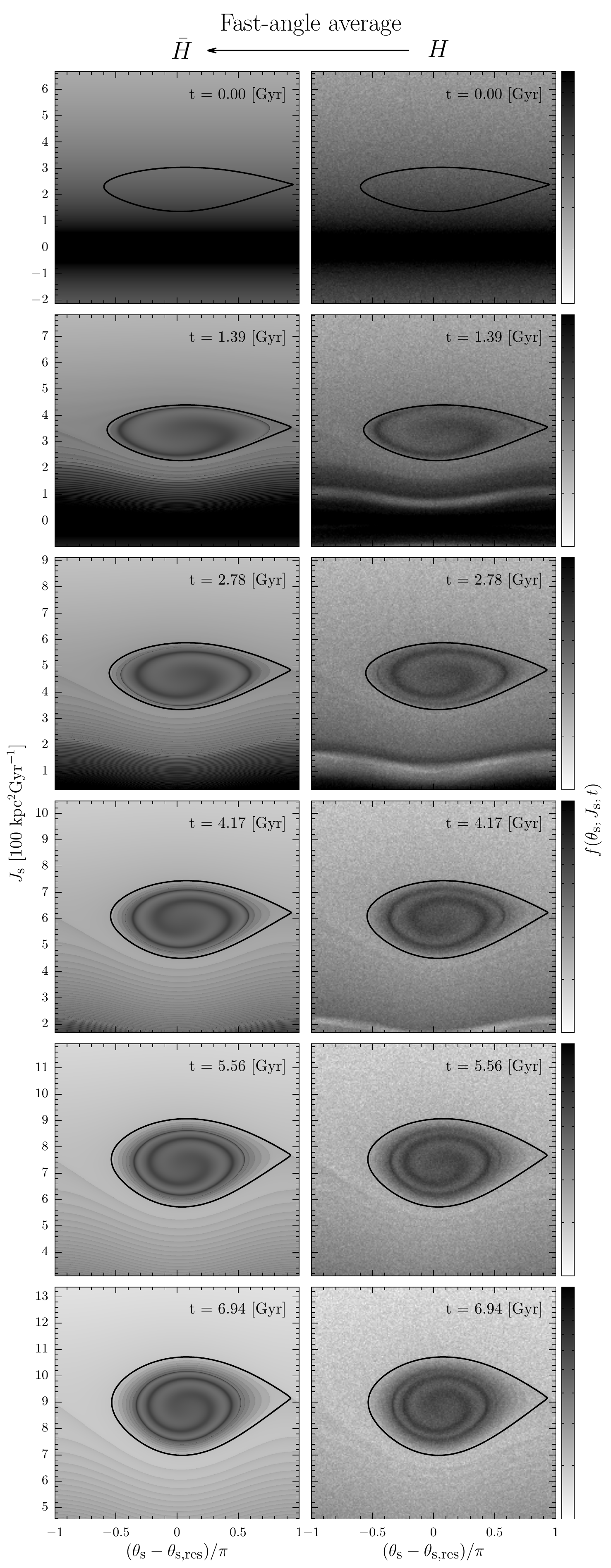}
    \caption{Time evolution of the phase-space distribution around a moving resonance in the slow regime ($\eta = 0.002, s \sim 0.16$). Left column: evolution with the 1D averaged Hamiltonian $\bH$ using the CBE (equation \ref{eq:CBE_sol_slow}). Right column: evolution with the 3D full Hamiltonian $H$ using a standard test-particle simulation.}
    \label{fig:f_eta0002}
  \end{center}
\end{figure}

In contrast to the study in the slow limit \citep{Chiba2022Oscillating}, where analytical solutions to the equations of motion were available, here we solve the Hamilton's equations numerically following the method of \cite{Weinberg2007BarHaloInteraction}. This allows us to abandon the pendulum approximation, making the averaging in $\vthetaf$ the only approximation we rely on. Note that the result of the numerical calculation does not hinge on whether we use the static or comoving action coordinate. For the sake of computational simplicity, we use the former. The latter will be used to demarcate the phase space of trapping and to model \textit{dynamical feedback} in Section \ref{sec:dynamical_feedback}. We summarize the computational method in Appendix \ref{sec:numerical_method}.

Figure~\ref{fig:f_eta0002} shows the time-evolution of the phase-space density in the slow angle-action plane $(\thetas,\Js)$. As before, we choose the corotation resonance $\vN=(0,2,2)$ at $(\Jfo,\Jft)=(10,0) \kpc^2 \Gyr^{-1}$. The bar forms at $t=0$ with pattern speed $\Omegap=72\Gyr^{-1}$ and spins down to $36\Gyr^{-1}$ in approximately $6.9 \Gyr$ with slowing rate $\eta=0.002$. This results in the speed parameter increasing from $s \simeq 0.134$ to $0.184$, which falls under the slow regime ($s<1$).

The left column of Fig.~\ref{fig:f_eta0002} shows the DF calculated using the 1D averaged Hamiltonian $\bH$. The black closed curves mark the separatrix of the resonance defined by the time-frozen Hamiltonian in the comoving frame $\hH$ (equation \ref{eq:H_averaged_comoving}). The resonance is initially at small $\Js$, where the phase-space density is high. As the bar slows, the resonance moves towards larger $\Js$, carrying the trapped orbits and capturing new orbits as it grows in volume. The decrease in background density may cause an optical illusion that the trapped region is getting denser, but in fact the phase-space density of the trapped region is conserved, as ensured by the CBE.

The right column of Fig.~\ref{fig:f_eta0002} shows the snapshots obtained from a standard test-particle simulation using the 3D full Hamiltonian $H$. We integrated orbits in the galactic plane ($\Jft = 0$) forward in time and measured the density in the range $\Jfo \in [0,20] \kpc^2 \Gyr^{-1}$. The 1D averaged model describes the full 3D simulation remarkably well. The only structure the 1D model fails to reproduce is the groove seen at $\Js \approx 100 \kpc^2 \Gyr^{-1}$ which is created by the inner 1:4 resonance. This is expected since averaging the Hamiltonian over the fast angles $\vthetaf$ effectively removes all Fourier series of perturbations other than the target resonance.

\subsection{Friction and feedback in the slow-nonlinear regime}
\label{sec:friction_feedback_slow_regime}

Figure~\ref{fig:f_eta0002} clarifies the mechanism by which angular momentum is transferred from the bar to the halo in the general slow regime. We describe this based on two classifications: (i) trapped or untrapped, and (ii) dynamical friction or dynamical feedback.

\subsubsection{Dynamical friction by trapped orbits}
\label{subsec:friction_trapped}

In the trapped region, the motion of libration generates a winding phase-space spiral, which gives rise to an oscillating dynamical friction on the perturber. This was the main point of discussion in \cite{Chiba2022Oscillating} and \cite{Banik2022Nonperturbative}. The phase spiral emerges because the libration frequency drops towards the separatrix and the density along the motion of libration is non-uniform due to the initial negative gradient in $f$. Phase mixing acts to erase this gradient and, during this process, net angular momentum is transferred to the halo. As mixing proceeds, the frictional torque undergoes damped oscillations. However, because the libration region is growing in size and capturing new orbits, fresh inhomogeneity is constantly introduced into the trapped region, preventing dynamical friction from vanishing entirely.

\subsubsection{Dynamical feedback by trapped orbits}
\label{subsec:feedback_trapped}

When the bar spins down due to dynamical friction, the resonance accordingly moves, typically towards larger angular momentum (Appendix \ref{sec:G_Gamma}). Since the trapped orbits comove with the resonance, they gain angular momentum, meaning that an additional negative torque is applied on the bar. Following \citetalias{Tremaine1984Dynamical}, we refer to this type of torque as \textit{dynamical feedback}, although the original definition differs (see Sections \ref{subsec:feedback_untrapped} and \ref{sec:dynamical_feedback}). Dynamical feedback is fundamentally different from friction in that it arises only when the frequency of the perturber changes. Since dynamical feedback is proportional to the rate at which the bar slows down, it modifies the bar's effective moment of inertia, typically reducing it. This allows the bar to spin down more rapidly in response to dynamical friction. We will formulate and quantify dynamical feedback in Section \ref{sec:dynamical_feedback}.

\subsubsection{Dynamical friction by untrapped orbits}
\label{subsec:friction_untrapped}

The untrapped orbits circulating on either side of the resonance also exert dynamical friction in much the same manner as the trapped orbits: the motion of circulation mixes orbits with different initial $\Js$, and the negative gradient in $f_0$ assures that this results in a net transfer of angular momentum to the halo. This dynamical friction works regardless of the value of $s$.

\subsubsection{Dynamical feedback by untrapped orbits}
\label{subsec:feedback_untrapped}

In the slow regime, untrapped orbits transiting from one side of the resonance to the other experience a net change (typically a loss) in angular momentum that is proportional to the volume of the trapped phase-space. This torque is caused only when the resonance shifts in position and is therefore classified as dynamical feedback. In fact, it was this feedback torque by the untrapped orbits that \citetalias{Tremaine1984Dynamical} first identified and dubbed `dynamical feedback'. As with the feedback from trapped orbits, this torque acts to modify the bar's moment of inertia, albeit in the opposite direction. The total dynamical feedback is given by the difference between the two and is typically negative, hence promoting the bar's spin-down (see Section \ref{sec:dynamical_feedback}).

\subsection{Wave form of the perturbation}
\label{sec:wave_form}

A close inspection of Fig.~\ref{fig:f_eta0002} reveals that the phase spirals in the trapped region have two different wave forms: (i) At the core of the resonance, the phase spiral is continuous, as it stems from the smooth initial distribution of orbits captured instantly at the time of bar formation. (ii) At the outer region, which formed later, the density varies discontinuously between the neighbouring layers. This discontinuity arises from orbital capture at the saddle point of the Hamiltonian (cf. Figs.~\ref{fig:H_ModPend},\ref{fig:schematic_effective_potential}). At this point, two separate phase-space trajectories approach from above and below and become adjacent upon entering the trapped region. Since these newly trapped orbits were not adjacent prior to capture, the phase-space density in the outer trapped region becomes discontinuous.

Similarly, in the untrapped region, the phase stripes\footnote{It should be noted that phase `spirals' and `stripes' are coordinate-dependent patterns: the spirals in the trapped phase-space appear as stripes if plotted in, e.g., the angle-action of libration $(\thetal,\Jl)$, while the stripes in the untrapped phase-space appear as spirals if plotted in, e.g., Cartesian canonical coordinates $(\sqrt{2\Js}\cos\thetas,\sqrt{2\Js}\sin\thetas)$.} exhibit two distinct wave forms: (i) The stripes below the initial position of the resonance $(\Js \lesssim 200 \kpc^2 \Gyr^{-1})$ appear smooth as they originate from the instantaneous formation of the bar. (ii) In contrast, perturbations left by the passage of the resonance have a discontinuous form again due to the saddle point of $\hH$.

\begin{figure*}
  \begin{center}
      \includegraphics[width=17.8cm]{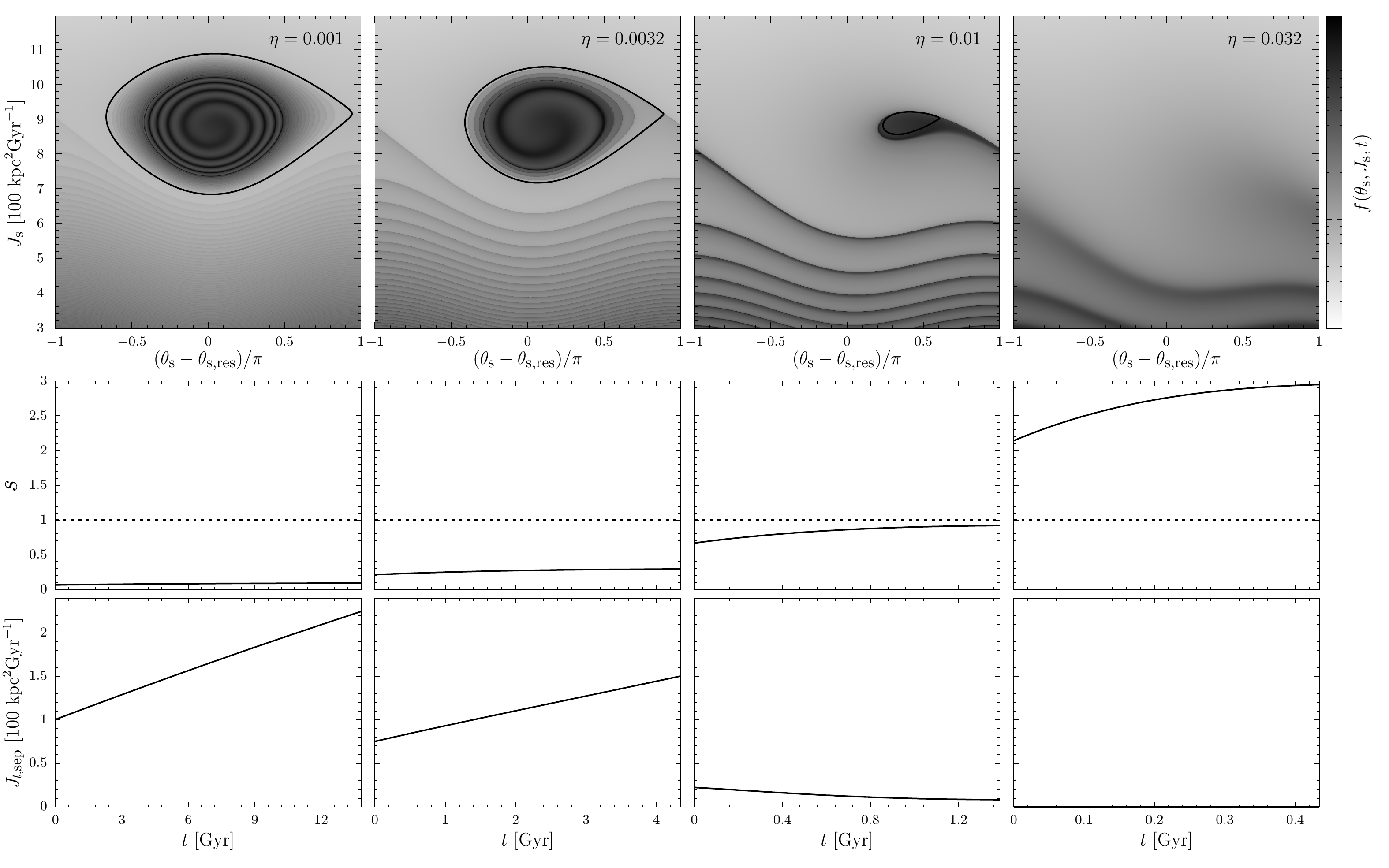}
    \caption{Dependence of the phase-space distribution near a resonance on the bar's slowing rate $\eta$. Top row: the distribution after the bar has slowed from $\Omegap=72$ to $36 \Gyr^{-1}$. Middle row: time evolution of the speed parameter $s$ (equation \ref{eq:speed_parameter}). The dotted lines mark the critical value $s=1$ above which trapping is absent. Bottom row: time evolution of the phase-space area of trapping represented by $\Jlsep$ (equation \ref{eq:Jlsep}). Decrease in $\Jlsep$ at $\eta=0.01$ (third column) implies resonant escape.}
    \label{fig:f_eta}
  \end{center}
\end{figure*}

\subsection{Variation with bar slowing rate}
\label{sec:variation_with_bar_slowing_rate}

Since our method is applicable to any speed $s$, it allows us to seamlessly connect the slow and fast limits previously explored via adiabatic theory and linear theory, respectively. Figure~\ref{fig:f_eta} shows the variation of the phase-space distribution with slowing rate $\eta$. We chose $\eta$ such that they are approximately evenly spaced in logarithm with width $\Delta \log_{10} \eta \approx 0.5$. The top row plots the distribution after the bar slowed from $\Omegap=72$ to $36\Gyr^{-1}$. The left-three figures show evolution in the slow regime, where the speed parameter $s$ remains below unity as reported in the middle row. As described in Section \ref{sec:phase_space_dynamics}, the size of the trapped region diminishes with increasing $\eta$. For $\eta=0.001$ and 0.0032, we observe new trapped layers because the trapped volume increases with time, as shown in the bottom row. For $\eta=0.01$, however, the volume reduces, causing trapped orbits to escape the resonance and create a dense trailing wake.

\begin{figure*}
  \begin{center}
      \includegraphics[width=17.8cm]{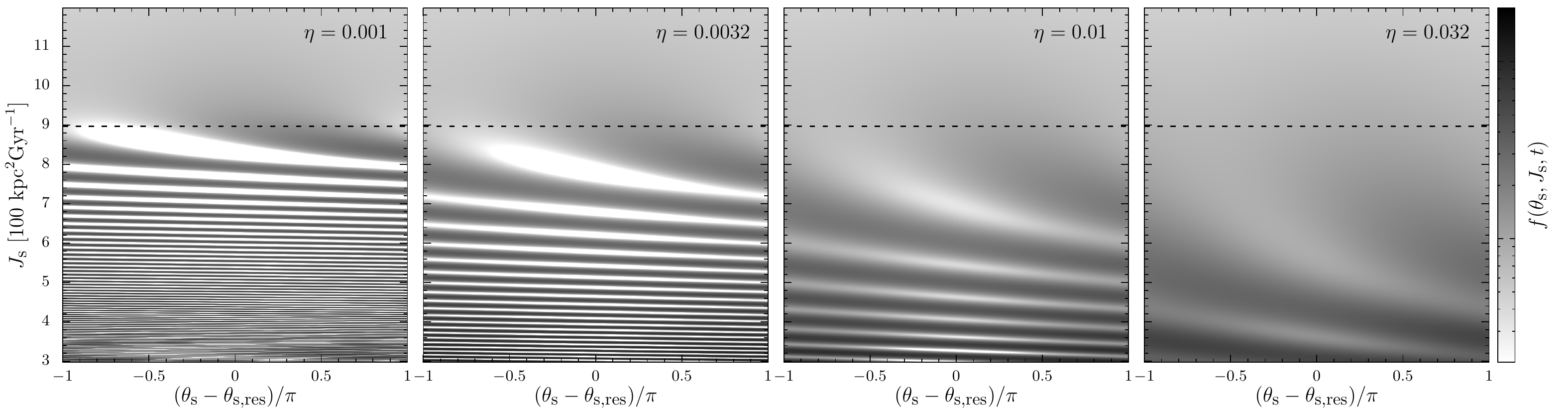}
    \caption{Phase-space distribution near a moving resonance predicted by linear perturbation theory (equation \ref{eq:f1_linearizedCBE_averaged_3}). Dotted lines mark the current position of the resonance. Linear theory qualitatively describes the near-resonant dynamics (Fig.~\ref{fig:f_eta}) only in the fast regime (right figure) where libration is absent.}
    \label{fig:f_eta_linear}
  \end{center}
\end{figure*}

Whether the trapped volume increases or not depends on a number of factors, in particular, from equation (\ref{eq:Jlsep}),
\begin{align}
  \frac{\drm \Jlsep}{\drm t} = \frac{8J_0}{\pi}\left[\frac{1}{2}\left(\frac{\Psi'}{\Psi}-\frac{G'}{G}\right) C + \frac{\drm C}{\drm s} s' \right],
  \label{eq:dJlsepdt}
\end{align}
where the prime denotes a time derivative. Since our bar potential has a fixed amplitude and elongates along with the expanding resonance, $\Psi'/\Psi$ is minor. In contrast, $-G'/G$ is large and positive since $G$ rapidly approaches zero as the resonance moves towards larger $\Js$ (Appendix \ref{sec:G_Gamma}). The third term on the r.h.s of equation (\ref{eq:dJlsepdt}) describes the change in the resonant-volume factor $C$ (equation \ref{eq:resonant_volume_factor}) and is typically negative since $\drm C/ \drm s<0$ while $s'>0$ (Fig.~\ref{fig:f_eta} middle row). For small $s$, and thus $C \sim 1$, the second term dominates, resulting in an increasing trapped volume. However, in the transition regime (left-third column of Fig.~\ref{fig:f_eta}), where $s$ is close to unity and thus $C$ is small (Fig.\ref{fig:resonant_volume_factor}), the third term becomes dominant, causing the resonance to shrink.

The right column of Fig.~\ref{fig:f_eta} shows evolution in the fast regime where trapping is absent. The striated perturbation remains, but it now has a smooth distribution since neighbouring orbits no longer split at the resonance due to the disappearance of the saddle points of $\hH$ (cf. Fig.~\ref{fig:H_ModPend}).

\subsection{Comparison with linear perturbation theory}
\label{sec:comparison_linear_theory}

The perturbed distribution in the fast regime should align with predictions from linear perturbation theory since librating orbits are absent. To confirm this, we derive the linear response in the slow angle-action space, allowing for the general time-dependence of the perturber. The linearized CBE has the following general solution \citep[][]{Weinberg2004Timedependent}:
\begin{align}
  f_1(\vtheta, \vJ, t)
  &= \sum_\vn \left[i \vn \!\cdot\! \frac{\pd f_0}{\pd \vJ} \int_0^t \drm t' \e^{- i \vn \cdot \vOmega (t - t')} \whPhin(\vJ, t') \right] \e^{i \vn \cdot \vtheta}.
  \label{eq:f1_linearizedCBE_1}
\end{align}
We separate the rotational time-dependence of the perturbation $\whPhin(\vJ,t)=\Phin(\vJ,t)\e^{-i\nphi \int^t_0 \drm t' \Omegap(t')}$ and rewrite
\begin{align}
  f_1(\vtheta, \vJ, t)
  &= i \sum_\vn \vn \!\cdot\! \frac{\pd f_0}{\pd \vJ} \e^{i \left[\vn \cdot \vtheta -\nphi \int^t_0 \drm t' \Omegap(t')\right]} \nonumber \\
  &\hspace{3mm}\times \int_0^t \drm t' \e^{- i \int_{t'}^t \drm t'' \Omegas(\vJ,t'')} \Phin(\vJ, t'),
  \label{eq:f1_linearizedCBE_2}
\end{align}
where $\Omegas(\vJ,t) \equiv \vn \cdot \vOmega(\vJ) - \nphi \Omegap(t)$. We transform $\vtheta$ to the fast-slow angles and average $f_1$ over the fast angles
\begin{align}
  \bar{f}_1(\thetas, \vJ, t)
  &= \frac{1}{(2\pi)^2} \int \drm^2 \vthetaf ~i \sum_\vn \vn \!\cdot\! \frac{\pd f_0}{\pd \vJ} \nonumber \\
  &\hspace{3mm}\times \e^{i \left[\left(\nr - \Nr \frac{\nphi}{\Nphi}\right)\thetafo + \left(\npsi - \Npsi \frac{\nphi}{\Nphi}\right)\thetaft + \frac{\nphi}{\Nphi}\thetas \right]} \nonumber \\
  &\hspace{3mm}\times \int_0^t \drm t' \e^{- i \int_{t'}^t \drm t'' \Omegas(\vJ,t'')} \Phin(\vJ, t').
  \label{eq:f1_linearizedCBE_averaged_1}
\end{align}
Since $\Phin$ is non-zero only when $\nphi=\pm m = \pm 2$ \citep[][Appendix B]{Chiba2022Oscillating}, we have for resonances with $\Nphi = 2$,
\begin{align}
  \bar{f}_1(\thetas, \vJ, t)
  &= i \sum_\vn \vn \!\cdot\! \frac{\pd f_0}{\pd \vJ} \delta_{\nr,\pm\Nr} \delta_{\npsi,\pm\Npsi} \delta_{\nphi,\pm\Nphi} \nonumber \\
  &\hspace{3mm}\times \int_0^t \drm t' \e^{i\left[\frac{\nphi}{\Nphi}\thetas - \int_{t'}^t \drm t'' \Omegas(\vJ,t'')\right]} \Phin(\vJ, t'),
  \label{eq:f1_linearizedCBE_averaged_2}
\end{align}
where $\delta_{n,N}$ is the Kronecker delta. Using the reality condition $\Phi_{-\vn} = \Phin^{*}$ and the relation $\vN \!\cdot \left( \pd f_0/\pd \vJ \right) = \pd f_0/\pd \Js$, we finally obtain
\begin{align}
  \bar{f}_1(\thetas, \vJ, t) &= - \frac{\pd f_0}{\pd \Js} \int_0^t \!\drm t' \sin\left[\thetas - \thetasres - \int_{t'}^t \!\drm t'' \Omegas(\vJ,t'') \right] \nonumber \\
  &\hspace{3mm}\times |\PhiN(\vJ,t')|.
  \label{eq:f1_linearizedCBE_averaged_3}
\end{align}
Figure~\ref{fig:f_eta_linear} plots the perturbed distribution $f=f_0+\bar{f}_1$ predicted by equation (\ref{eq:f1_linearizedCBE_averaged_3}). Comparison with Fig.~\ref{fig:f_eta} confirms that linear theory is qualitatively correct in the fast regime (right figure), although the errors are catastrophic in the slow regime (left three figures) where resonant orbits librate. Note that the accuracy of linear theory also depends on the evolution time, which shortens as the slowing rate increases. We have confirmed, however, that linear theory continues to succeed in the fast regime even beyond the typical libration period $t_{\rm lib}$ since the nonlinear response is precluded \citepalias{Tremaine1984Dynamical}. Hence, while linear theory works in the slow regime ($s<1$) only for a short period ($t \lesssim t_{\rm lib}$), it remains valid in the fast regime ($s>1$) over an extended period ($t \gtrsim t_{\rm lib}$).

\section{Density response}
\label{sec:density_response}

Although the phase-space distribution contains the full information of the halo's response, the density response in real space will help us understand the nature of dynamical friction and feedback.

We calculate the density response by integrating the distribution function over a Cartesian grid ($96\times 96 \times 96$) in 3-dimensional velocity space:
\begin{align}
  \rho(\vx,t) = \int \drm^3 \vvel f(\thetas,\vJ,t).
  \label{eq:rho}
\end{align}
For each grid point, we compute $f$ by transforming the coordinates to the fast-slow angle-action variables and evolving only the slow variables backward in time until $t=0$ (equation \ref{eq:CBE_sol_slow}). In practice, we truncate the velocity integral at a fixed energy $E_{\rm max} = \Phi_0(r_{\rm max})$, where we set $r_{\rm max} = 10 \rs = 200 \kpc$. Calculations with $r_{\rm max} > 10 \rs$ made little visible difference.

\begin{figure*}
  \begin{center}
    \includegraphics[width=17.cm]{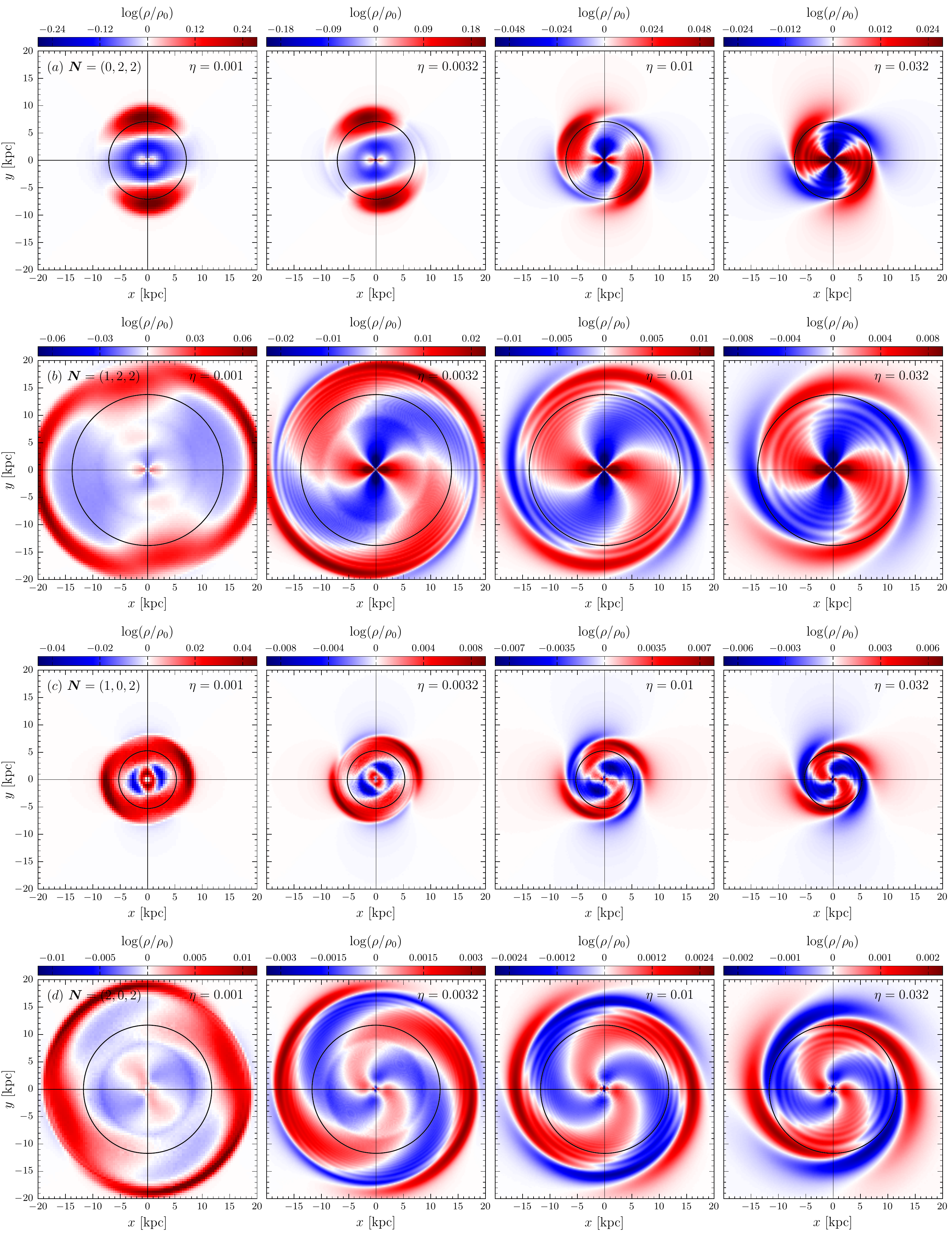}
    \caption{Density wakes at the equatorial plane ($z=0$) caused by (a) the corotation resonance $\vN=(0,2,2)$, (b) the outer Lindblad resonance $\vN=(1,2,2)$, and (c,d) the direct radial resonances $\vN=(1,0,2)$ and $\vN=(2,0,2)$. The density is calculated from the distribution function evolved with the fast-angle averaged Hamiltonian. The bar lies on the $x$-axis and is rotating anti-clockwise. The black circles mark the resonant radius for circular orbits ($\Jr=0$).}
    \label{fig:dens_individual}
  \end{center}
\end{figure*}

\begin{figure*}
  \begin{center}
    \includegraphics[width=17.5cm]{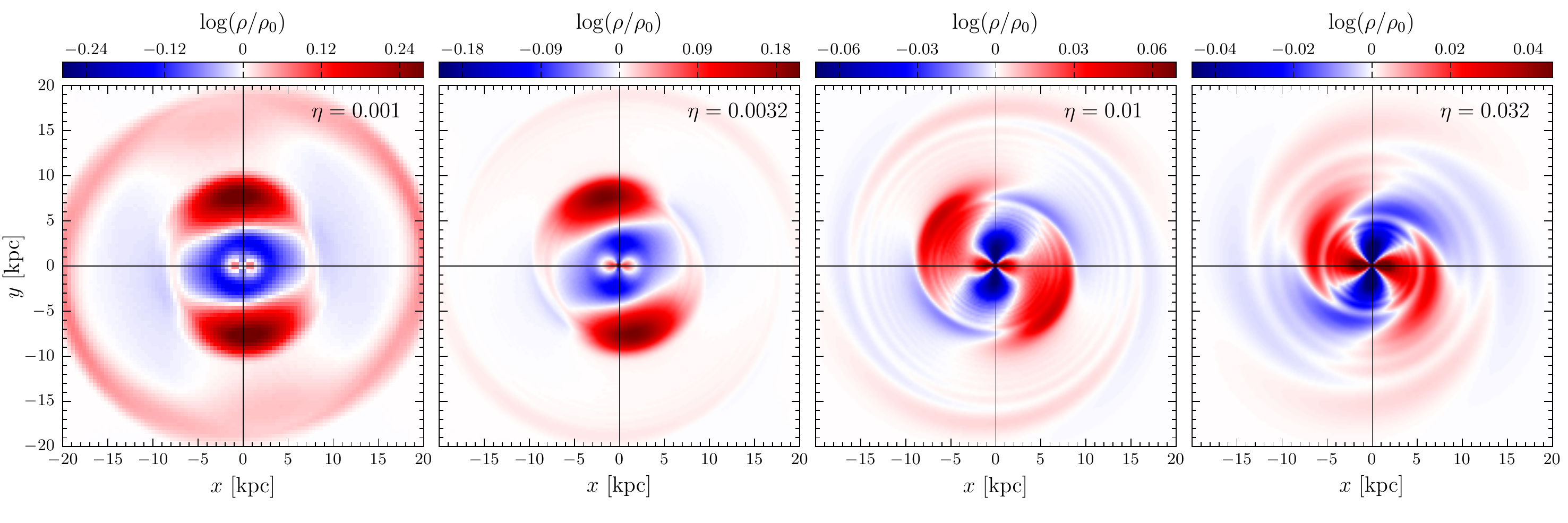}
    (a)~Superposition of density wakes generated by the four dominant resonances using the 1D averaged Hamiltonian.
    \includegraphics[width=17.5cm]{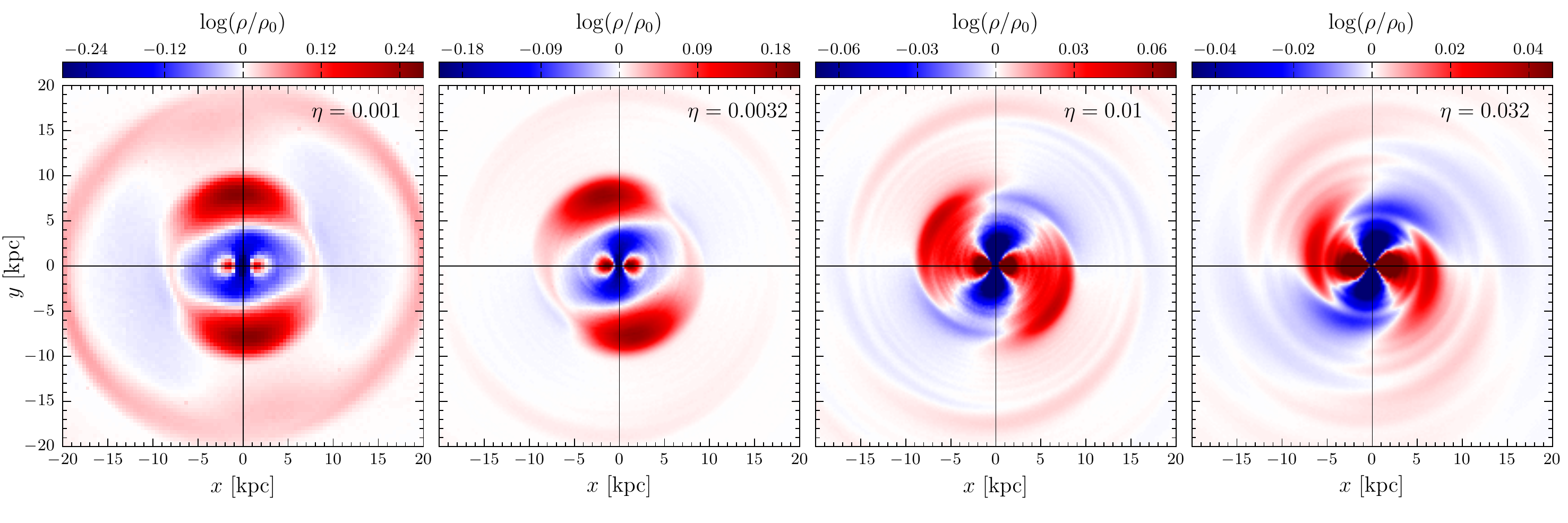}
    (b)~Standard test-particle simulation using the 3D Hamiltonian in Cartesian coordinate.
    \includegraphics[width=17.5cm]{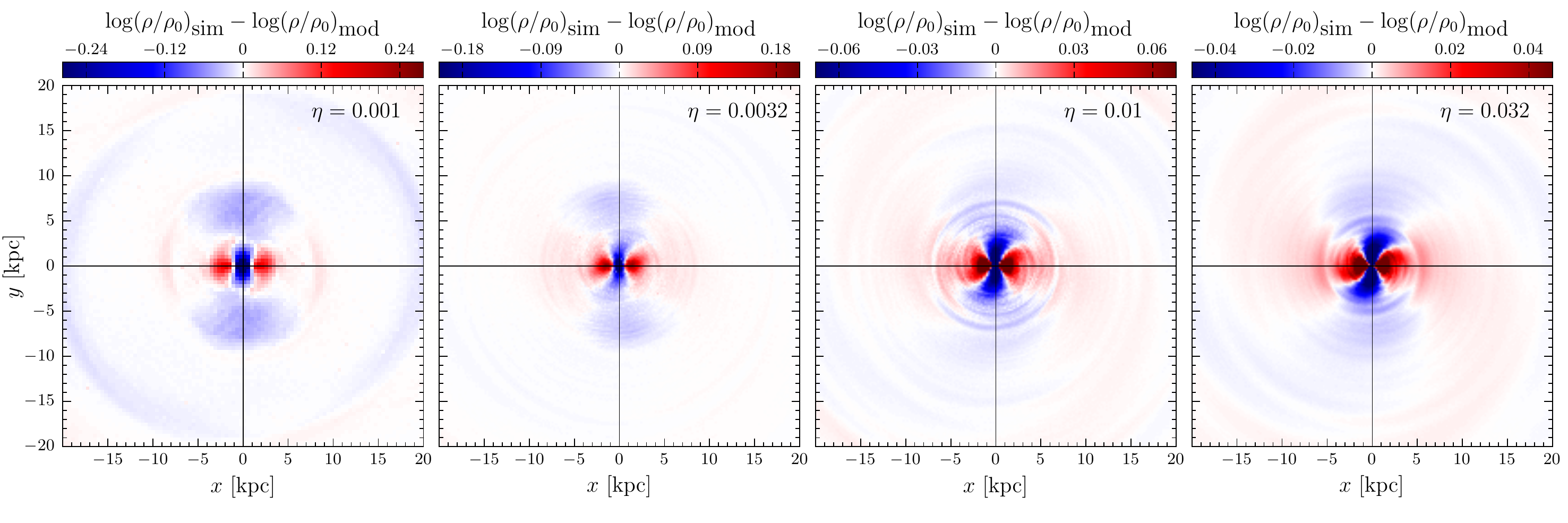}
    (c)~Residual between the 1D model and the 3D simulation.
    \caption{Density wake at the galactic plane calculated using the 1D averaged Hamiltonian (top) and the 3D full Hamiltonian (middle). Bottom row shows the residual between the two.}
    \label{fig:dens_averH_fullH}
  \end{center}
\end{figure*}

Figure \ref{fig:dens_individual} shows the density response at the galactic plane ($z=0$) caused by the four strongest resonances: the corotation resonance $\vN=(0,2,2)$, the outer Lindblad resonance $\vN=(1,2,2)$, and the direct radial resonances $\vN=(1,0,2)$ and $\vN=(2,0,2)$. The bar lies on the $x$-axis and is rotating anticlockwise. As before, we slow the bar from $\Omegap=72$ to $36\Gyr^{-1}$.

The density response varies significantly with slowing rate $\eta$ as the evolutionary regime makes a transition from slow to fast. For $\eta$ as small as $0.001$ (Fig.~\ref{fig:dens_individual} left column), the resonances mostly evolve in the slow regime, so the trapped orbits dragged from the inner halo form a \textit{density island} migrating outwards with the resonance. These density islands are misaligned with the bar because the center of libration $\thetares$ shifts anti-clockwise when the resonance is moving (see discussion surrounding Figs.~\ref{fig:H_ModPend}-\ref{fig:resonant_volume_factor}). As $\eta$ is increased from left to right, the density island rotates further. Eventually, the evolution shifts to the fast regime, and the density response deforms into a striated pattern -- a \textit{density wake} -- that no longer drifts with the resonance. Note that slow and fast regimes typically coexist since orbits crossing the galactic plane take a wide range of $\vJf$ and thus pass the resonance at different speed $s$.

It is worth noting that the density response in the fast/linear regime is mostly \textit{in phase} with the bar, implying that the halo's self-gravity will amplify the bar's perturbation \citep[][]{Dootson2022}. However, the density response gradually becomes \textit{out of phase} with the bar as the evolution becomes slow/nonlinear. This implies that, in the slow regime, the self-gravity of the halo may weaken the bar's perturbation, causing the bar to evolve even slower.

Figure \ref{fig:dens_averH_fullH} compares the superposition of the density response induced by the four strongest resonances (top row) with the 3D test-particle simulation (middle row). The superposition of 1D averaged models reproduces almost all features of the 3D simulation. This result confirms that the density response for each resonance has been computed correctly and that the assumption behind the method of averaging (i.e. resonances are well separated) is acceptable. The bottom row shows the residual between the 1D model and the 3D simulation. At small $\eta$ (left column), the residual shows weakly negative (blue) features in the trapped region which indicate that the trapped mass in the full simulation is slightly smaller than in the averaged model. This is likely due to the partial overlap of resonances \citep[Fig.8 and Appendix F of][]{Chiba2022Oscillating}, where the chaotic behavior of orbits can hinder them from remaining trapped on the moving resonances. The residual is also non-negligible near the galactic centre, where all resonances induce coherent quadrupole perturbations, implying that adding contributions from other minor resonances may further improve the model.

\section{Dynamical feedback}
\label{sec:dynamical_feedback}

In this section, we formulate dynamical feedback introduced by \citetalias{Tremaine1984Dynamical}. As described in Section \ref{subsec:feedback_trapped}, dynamical feedback is the transfer of momentum that arises directly from the migration of the trapped phase-space. Suppose, for instance, that the bar slowed down by dynamical friction and, as a result, the resonance shifted towards larger angular momentum. The orbits trapped in resonance will then be dragged along with the resonance, carrying further angular momentum away from the bar. Dynamical feedback refers to this feedback torque. Dynamical feedback only occurs in the slow nonlinear regime, where there is a finite volume of trapping, and is thus not predicted by standard linear perturbation theory \citep[e.g.][]{lynden1972generating,Weinberg2004Timedependent}.

Dynamical feedback directly depends on the rate of change of the bar's pattern speed which has an interesting consequence. To see this, let us write dynamical feedback as $\taufb = \Ifb \dOmegap$, where $\Ifb$ is a coefficient that has the dimension of a moment of inertia, and denote the usual dynamical friction as $\tau_{\rm fr}$. The equation of motion of the bar $I \dOmegap = \tau_{\rm fr} + \tau_{\rm fb}$ (Section \ref{sec:model}) can then be written as 
\begin{align}
  (I - \Ifb) \dOmegap = \tau_{\rm fr},
  \label{eq:change_in_Ib}
\end{align}
which implies that dynamical feedback changes the moment of inertia of the perturber \citep{weinberg1985evolution}. If $\Ifb>0$, dynamical feedback helps the bar spin down (a positive feedback); otherwise, trapping stabilizes the bar's pattern speed (a negative feedback). As we shall see, dynamical feedback by the trapped orbits is typically positive.

The original study by \citetalias{Tremaine1984Dynamical} in fact did not concern dynamical feedback by the trapped orbits. Instead, they studied feedback from the \textit{untrapped} orbits passed by a slowly moving resonance. The feedbacks from trapped and untrapped orbits are in fact closely related through the collisionless Boltzmann equation, which states that phase-space flow is incompressible: When the trapped island moves, the untrapped orbits in its path are forced to skirt around it and flow to the far side of the resonance, as illustrated in Fig.~\ref{fig:H_ModPend}. Therefore, an acceleration (deceleration) of the trapped orbits must be accompanied by a deceleration (acceleration) of the surrounding untrapped orbits. The net dynamical feedback is determined by the imbalance between these two effects.

Here, we provide the full formula of dynamical feedback, incorporating contributions from both trapped and untrapped orbits.

\subsection{Dynamical feedback by trapped orbits}
\label{subsec:dynamical_feedback_trapped}

We begin with the dynamical feedback by the trapped orbits. Since dynamical feedback is the torque that arises purely from the movement of the trapped phase-space, we must first remove from the distribution function any transient perturbations that phase mixes over time. This is effected by averaging the distribution function over the angle of libration
\begin{align}
  \flmix(\vJf,\Jl,t) = \frac{1}{2\pi} \int \drm \thetal \fl(\thetal,\vJf,\Jl,t).
  \label{eq:f_phase_mixed}
\end{align}
We multiply $\flmix$ with minus the torque on individual halo orbits
\begin{align}
  \dLz = \Nphi \left(\dJsres + \dDelta\right)
  \label{eq:dLzdt}
\end{align}
and integrate over the slow phase-space $\drm\thetas\drm\Js=\drm\thetal\drm\Jl$
\begin{align}
  \htaul\left(\vJf,t\right) = - \Nphi \int_0^{\Jlsep} \! \drm \Jl \int_{0}^{2\pi} \! \drm \thetal \flmix\left(\vJf,\Jl,t\right) \left( \dJsres + \dDelta \right).
  \label{eq:tauN_1}
\end{align}
This is the torque felt by the bar due to orbits trapped in resonance $\vN$ at a specific $\vJf$. Since $\dDelta$ is $2\pi$ periodic in $\thetal$, the second term vanishes, leaving
\begin{align}
  \htaul\left(\vJf,t\right) = - \frac{2\pi \Nphi^2 \dOmegap}{G(\vJf,\Jsres)} \int_0^{\Jlsep} \! \drm \Jl \flmix\left(\vJf,\Jl,t\right),
  \label{eq:tauN_2}
\end{align}
where we have plugged in equation (\ref{eq:dJsresdt}). We remind the reader that $G$ is $\pd (\vN\cdot\vOmega)/\pd \Js$ evaluated at the resonance $\Js=\Jsres$ (equation \ref{eq:GF}), and is not the gravitational constant.

Depending on the geometry of the resonance, there could exist more than one trapped region at a fixed $\vJf$ satisfying the \textit{same} resonance condition: the resonance curve on the $(L,\Jr)$ plane may have multiple intersections with the planes of constant $\vJf$. Such cases are limited but found, for instance, at the corotation resonance at small $L$ and large $\Jr$ (Appendix \ref{sec:G_Gamma}). Hence we write
\begin{align}
  \htaul\left(\vJf,t\right) = - 2\pi \Nphi^2 \dOmegap \sum_i \frac{\Bl(\vJf,\Jsres^{i},t)}{G(\vJf,\Jsres^{i})},
  \label{eq:tauN_3}
\end{align}
where the function
\begin{align}
  \Bl(\vJf,\Jsres^{i},t) = \int_0^{\Jlsep} \! \drm \Jl \flmix\left(\vJf,\Jl,t\right)
  \label{eq:Bl}
\end{align}
quantifies the amount of trapped orbits near each resonance $\Jsres^{i}$. Equation (\ref{eq:tauN_3}) can be formally rewritten as 
\begin{align}
  \htaul\left(\vJf,t\right) = - 2\pi \Nphi^2 \dOmegap \!\int\! \drm \Js \sgn(G) \Bl(\vJf,\Js,t) \delta\!\left(\vN \!\cdot\! \vOmega - \Nphi \Omegap \right),
  \label{eq:tauN_4}
\end{align}
using the identity $\sum_i f(x_i)/|g'(x_i)|=\int \drm x f(x) \delta(g(x))$ \citep[e.g.][]{kanwal1998generalized}, where $\delta$ is the Dirac's delta function.

To get the total dynamical feedback by trapped orbits, we integrate $\htaul$ over the fast angle-actions and sum over all resonances
\begin{align}
  \taul 
  &= \sum_{\vN} \int \drm^2 \vJf ~\drm^2 \vthetaf ~\htaul\left(\vJf,t\right) \nonumber \\
  &= - (2\pi)^3 \dOmegap \!\sum_{\vN} \Nphi^2 \! \int \! \drm^3 \vJ' \sgn(G) \Bl(\vJ'\!,t) \delta\!\left(\vN \!\cdot\! \vOmega - \Nphi \Omegap \right),
  \label{eq:taufb_trapped_1}
\end{align}
where we used the notation $\vJ'=(\vJf,\Js)$. Switching variables from $\vJ'$ to $\vJ=(\Jr,L,\Lz)$, where the Jacobian determinant is $\left\vert\pd (\vJ')/\pd (\vJ)\right\vert = 1/\Nphi$, we finally obtain
\begin{align}
  \taul = - (2\pi)^3 \dOmegap \!\sum_{\vN} \Nphi \!\! \int \!\! \drm^3 \vJ \sgn(G) \Bl(\vJ,t) \delta\!\left(\vN \!\cdot\! \vOmega - \Nphi \Omegap \right).
  \label{eq:taufb_trapped_final}
\end{align}
This equation describes the dynamical feedback from all the trapped orbits dragged by the moving resonances: the function $\Bl$ measures the mass of trapped orbits, and the $\sgn(G)$ encodes the direction at which they move in $\Lz$ in response to the change in pattern speed. In the absence of resonance migration ($\dOmegap=0$), dynamical feedback vanishes as expected. Since $\Bl \geq 0$ by definition and $G<0$ for most cases (Appendix \ref{sec:G_Gamma}), trapped orbits generally provide a net positive feedback on the perturber, thereby decreasing the perturber's moment of inertia.

\begin{figure*}
  \begin{center}
    \includegraphics[width=17.5cm]{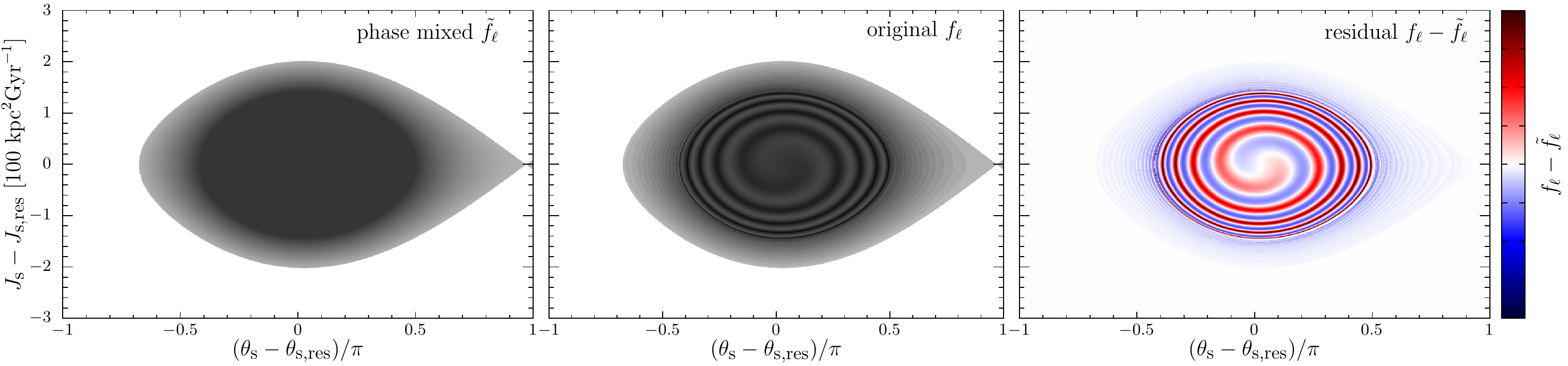}
    \caption{Left: model for the phase-mixed distribution function (equation \ref{eq:flmix_model}). Middle: original distribution function (equation \ref{eq:CBE_sol_slow}). Right: residual between the mixed model and full distribution. For all panels, the bar slowed from $\Omegap=72$ to $36 \Gyr^{-1}$ with slowing rate $\eta=0.001$.}
    \label{fig:f_mixed}
  \end{center}
\end{figure*}

\subsection{Dynamical feedback by untrapped orbits: the TW formula}
\label{subsec:dynamical_feedback_untrapped}

We next formulate dynamical feedback by the untrapped orbits. As in the previous section, we first aim at removing the transient perturbations from the distribution function by averaging it over the angle. Unlike the trapped orbits, however, the angle of circulation for the untrapped orbits cannot be defined from their trajectories in the comoving frame since the orbits are not closed (cf. Fig.~\ref{fig:H_ModPend}). The motion of circulation can instead be described by expanding perturbation series about the unperturbed orbit in the static frame. Perturbation theory then predicts that changes to the angle-averaged distribution function are of second order in the strength of perturbation $A$ \citep[e.g.][]{Fouvry2015WKBlimit}. Hence, to first order in $A$, the phase-mixed distribution of the untrapped (circulating) orbits $\fcmix$ can be approximated by the unperturbed distribution $f_0$. We will demonstrate below that this approximation leads to the formula derived by \citetalias{Tremaine1984Dynamical}.

With the aforementioned approximation, $\fcmix = f_0 + \mathcal{O}(A^2)$, the net torque on the bar by the untrapped halo orbits can be obtained by multiplying $f_0$ with $-\dLz$ and integrating it over the slow phase-space $\drm^2 S=\drm\thetas\drm\Js$, while excluding the region of libration $\Sl$ from the domain of integration:
\begin{align}
  \htauc\left(\vJf,t\right)
  =& - \int_{S-\Sl} \drm^2 S f_0(\vJf,\Js) \dLz \nonumber \\
  =& - \int_{S} \drm^2 S f_0(\vJf,\Js) \dLz + \int_{\Sl} \drm^2 S f_0(\vJf,\Js)\dLz.
  \label{eq:dynamical_feedback_NJf_start}
\end{align}
The first term in the second line is zero since the torque is periodic in the slow angle $\thetas$. We next Taylor expand $f_0$ in $\Js$ to first order around the resonance $\Jsres$
\begin{align}
  \htauc\left(\vJf,t\right) = \int_{\Sl} \drm^2 S \left[ f_0(\vJf,\Jsres) + \left.\frac{\pd f_0}{\pd \Js}\right\vert_{\Jsres} \Delta \right] \dLz,
  \label{eq:dynamical_feedback_NJf_expand}
\end{align}
where $\Delta = \Js - \Jsres$. We similarly expand the Hamiltonian (equation \ref{eq:H_averaged_comoving_Taylor}) which makes the second term vanish since the domain of integration $\Sl$ becomes symmetric about $\Delta=0$, and $\Delta$ is odd while $\dLz$ is even in $\Delta$. For the remaining first term, we can follow the same procedure as in the previous section beginning with equation (\ref{eq:tauN_1}). The dynamical feedback by all the untrapped orbits is then
\begin{align}
  \tauc = (2\pi)^3 \dOmegap \sum_{\vN} \Nphi \! \int \! \drm^3 \vJ \sgn(G) B_0(\vJ,t) \delta\!\left(\vN \cdot \vOmega - \Nphi \Omegap \right),
  \label{eq:taufb_untrapped_final}
\end{align}
where the function
\begin{align}
  B_0(\vJ,t) = \Jlsep(\vJ,t) f_0(\vJ)
  \label{eq:B0}
\end{align}
describes the mass of untrapped orbits displaced by the trapped orbits.

Substituting $B_0$ to equation (\ref{eq:taufb_untrapped_final}), and using equation (\ref{eq:Jlsep}), we obtain
\begin{align}
  \tauc
  = (2\pi)^3 \dOmegap \! \sum_{\vN} \Nphi \!\! \int \!\! \drm \vJ \sgn(G) J_0 Q(s) f_0(\vJ) \delta\!\left(\vN\!\cdot\!\vOmega\!-\!\Nphi\Omegap\right),
  \label{eq:dynamical_feedback_TW84}
\end{align}
where $Q(s)=8C(s)/\pi$ (equation \ref{eq:Q}) and $J_0=\sqrt{F/G}=\sqrt{\Psi/|G|}$ (equation \ref{eq:w0_J0}). Equation (\ref{eq:dynamical_feedback_TW84}) is the formula of dynamical feedback derived by \citetalias{Tremaine1984Dynamical} (equations 86,100) and summarized in \cite{weinberg1985evolution} (equation 39).

\subsection{Total dynamical feedback}
\label{subsec:dynamical_feedback_total}

The formulae derived in the previous sections can be combined to give the total dynamical feedback 
\begin{align}
  \taufb
  &= \taul + \tauc \nonumber \\
  &= (2\pi)^3 \dOmegap \sum_{\vN} \Nphi \! \int \! \drm^3 \vJ \sgn(G) \left(B_0 - \Bl\right) \delta\!\left(\vN \!\cdot\! \vOmega - \Nphi \Omegap \right).
  \label{eq:taufb_total}
\end{align}
This equation states that the strength of dynamical feedback is determined by the difference between the mass of the trapped orbits $\Bl$ and the mass of the background orbits $B_0$ that would otherwise occupy the trapped volume. This is a consequence of the incompressibility of phase space and is similar to the physics of buoyancy -- the trapped region is `submerged' in the phase space of untrapped orbits, and the torque required to move the trapped region is proportional to the density difference between the two regions.

The phase-space density of a dark halo in equilibrium typically declines with increasing angular momentum. Hence, so long as the resonance sweeps the phase space monotonically, the gainers of angular momentum always surpass the losers, resulting in a net negative torque on the bar. This negative torque provides a positive feedback when the bar decelerates while it provides a negative feedback when the bar accelerates. Thus dynamical feedback makes the bar easier to spin down than to spin up. If the bar switched between acceleration and deceleration in mid-course, dynamical feedback may provide a positive torque, resulting in the opposite conclusion. Generally, the stability condition for the bar's rotation depends on (i) the sign of the moment of inertia of the bar $I$, which was assumed positive in the discussion above, (ii) the sign of $G$ (equation \ref{eq:w0_J0}), and (iii) whether the trapped phase-space is overdense or not, $B_0-\Bl$. Thus
\begin{align}
  \begin{cases}
    \sgn (I) = \sgn \left[G\left(B_0 - \Bl\right)\right] ~~~ {\rm destabilizing ~(pos.~feedback)} \\
    \sgn (I) = -\sgn \left[G\left(B_0 - \Bl\right)\right] ~~~ {\rm stabilizing ~(neg.~feedback)}
  \end{cases}
  \label{eq:stability_condition}
\end{align}
Our stability condition differs from that of \citetalias{Tremaine1984Dynamical} (equation 104), which does not include feedback from trapped orbits, so $B_0 - \Bl$ is always positive. Since we expect $B_0 - \Bl < 0$ for a slowing bar, and $I>0$ while $G<0$ in most cases (Appendix \ref{sec:G_Gamma}), our revised stability condition implies that dynamical feedback will \textit{destabilize} the bar, as we demonstrate below.

To calculate equation (\ref{eq:taufb_total}), we need a model for the phase-mixed distribution function of the trapped region $\flmix$ (\ref{eq:f_phase_mixed}). For this, we extend the model of \cite{Sridhar1996Adiabatic} which provides a prescription for $\flmix$ in the extreme adiabatic limit $\eta/A \rightarrow 0$ (Section \ref{sec:adiabatic_theory}). In essence, the evolution of $\flmix(\Jl,t)$ is assumed adiabatic everywhere except at the separatrix. If the separatrix expands, orbits are captured into resonance, and the value of $\flmix$ at the expanding separatrix is given by the average density of the newly captured orbits. If the separatrix shrinks, orbits at the separatrix escape the resonance, so the domain of $\flmix$ simply diminishes but the distribution over the remaining domain is unaltered.

We extend this model to the adiabatic evolution in the comoving frame $\eta\gamma/A \rightarrow 0$ (Section \ref{sec:adiabatic_theory_comoving}) where the separatrix is deformed. In the extreme adiabatic limit, $\flmix$ at the expanding separatrix can be determined from the angle-averaged density of the untrapped orbits $\fcmix$ at the separatrices \citep[see][Table 1]{Sridhar1996Adiabatic}. However, as mentioned in Section \ref{subsec:dynamical_feedback_untrapped}, $\fcmix$ is not available in the comoving frame since untrapped orbits are not closed and thus the angles cannot be defined. As a crude approximation, we take $\flmix$ at the expanding separatrix to be the unperturbed density at the position of the resonance $\Jsres$ at the time of capture $t_{\rm cap}$:
\begin{align}
  \flmix(\vJf,\Jl) \simeq f_0(\vJf,\Jsres[t_{\rm cap}(\Jl)]).
  \label{eq:flmix_model}
\end{align}
Figure.~\ref{fig:f_mixed} compares this model (left) with the original distribution (middle) for $\eta=0.001$. Our model $\flmix$ predicts a flat distribution at the core because capture at the time of bar formation occurs while $\Jsres$ is fixed. The real phase-mixed distribution will not be flat unless $\pd f/\pd \Js$ is constant across the trapped region. Outside the core, our model $\flmix$ declines smoothly towards the surface. The residual between the model and the original $f$ (Fig.~\ref{fig:f_mixed}, right) exhibits the phase spiral with little offset, confirming that the model is reasonable.

\begin{figure}
  \begin{center}
    \includegraphics[width=8.5cm]{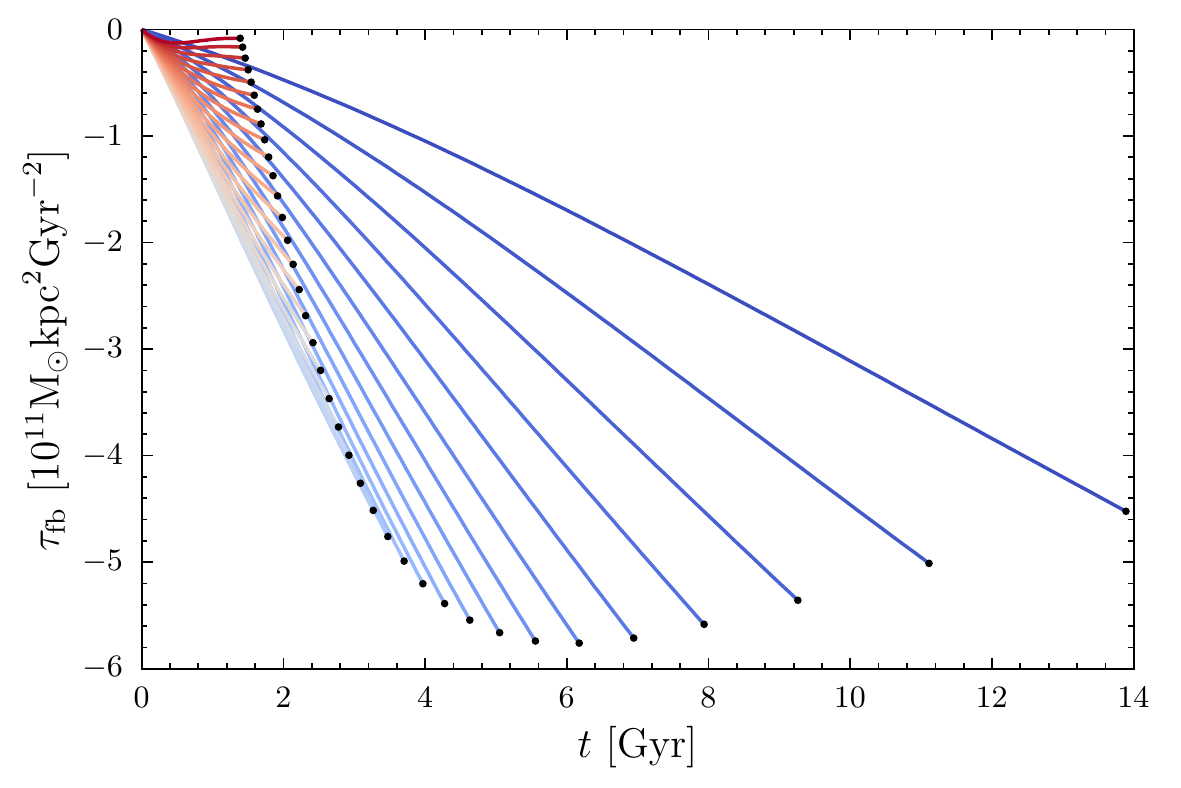}
    \caption{Time evolution of the dynamical feedback (\ref{eq:taufb_total}) by the corotation resonance. The slowing rate $\eta$ increases from 0.001 (blue) to 0.01 (red) with equal spacing 0.00025.}
    \label{fig:taufb}
  \end{center}
\end{figure}

Using this model for $\flmix$, we compute the dynamical feedback (\ref{eq:taufb_total}). Figure~\ref{fig:taufb} shows the dynamical feedback by the corotation resonance as a function of time for various slowing rate $\eta$, ranging from $0.001$ (blue) to $0.01$ (red). As before, we fix the bar's initial and final pattern speeds ($\Omegap=72$ to $36 \Gyr^{-1}$), so the time the bar takes to reach its final pattern speed (black dot) gets shorter with increasing $\eta$. The feedback torque is initially zero since $\Bl = B_0$, but decreases with time, becoming more negative, as the density contrast between the trapped and untrapped region gets higher (cf. Fig.~\ref{fig:f_eta0002}). Interestingly, feedback takes a peak at an intermediate slowing rate $\eta \sim 0.0025$.
\begin{figure}
  \begin{center}
    \includegraphics[width=8.5cm]{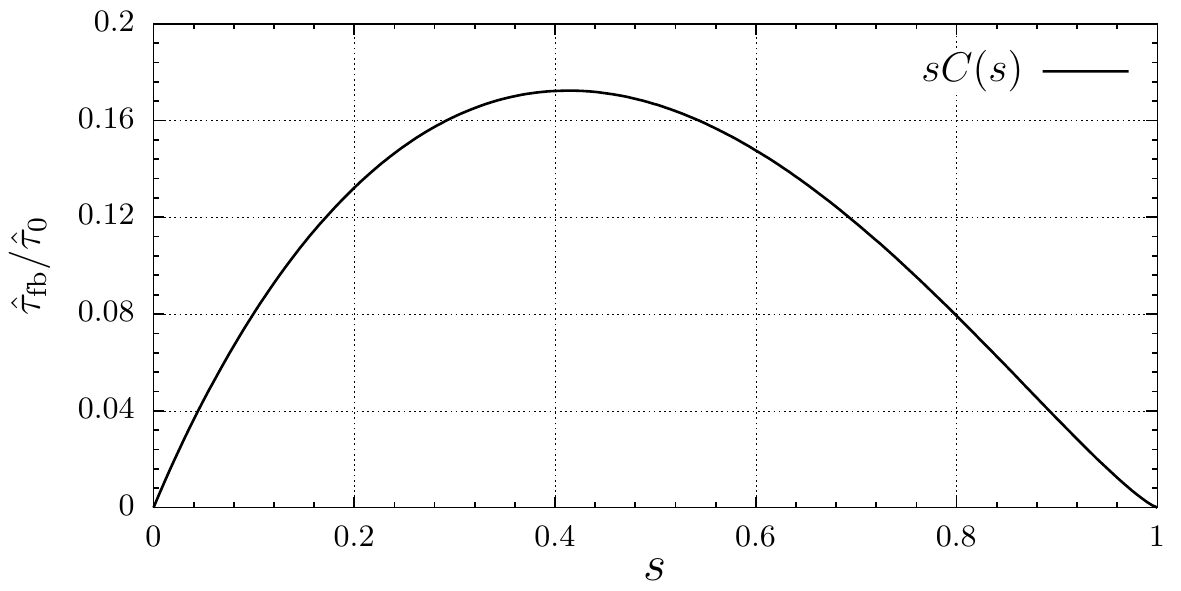}
    \caption{Dynamical feedback $\htau$ from orbits with a specific $\vJf$ as a function of the speed of the resonance $s$.}
    \label{fig:taufb_Jf}
  \end{center}
\end{figure}
To understand this, we analyse the feedback $\htaufb$ from orbits near a single point of resonance at a specific $\vJf$. From equation (\ref{eq:tauN_3}), we have
\begin{align}
  \htaufb
  &= \htaul + \htauc \nonumber \\ 
  &= 2\pi \Nphi^2 \dOmegap (B_0 - \Bl) / G = \sgn(\dOmegap G) \Nphi \Psi (B_0 - \Bl) s,
  \label{eq:dtau_1}
\end{align}
where $s = |\Nphi \dOmegap/ (G\Psi)|$ is the speed parameter (equation \ref{eq:speed_parameter}), allowing for arbitrary signs in $\dOmegap$ and $G$. Let us approximate $\Bl$ as $\Jlsep f_{\ell 0} = (8 J_0/\pi) C(s) f_{\ell 0}$, where $f_{\ell 0}$ is the averaged density of the librating region, and write
\begin{align}
  \htaufb = \htau_0 s C(s), ~~ {\rm where} ~~ \htau_0 = \sgn(\dOmegap G) \Nphi \Psi (8 J_0/\pi) (f_0 - f_{\ell 0}).
  \label{eq:dtau_2}
\end{align}
The equation indicates that feedback from each point on the resonance scales as $s C(s)$. Figure~\ref{fig:taufb_Jf} shows this torque $\htaufb$ as a function of $s$. At small speed $s$, the torque rises with increasing $s$ because the resonance shifts in angular momentum at a faster rate. As $s$ increases further, the torque saturates at $s \sim 0.4$ and starts declining because the trapped phase-space shrinks with $s$. The feedback drops to zero at $s=1$ where trapping vanishes entirely. The non-monotonic dependence of the total feedback on $\eta$ (Fig.~\ref{fig:taufb}) can be understood as the integration of this curve over the entire resonant surface.

\begin{figure}
  \begin{center}
    \includegraphics[width=8.5cm]{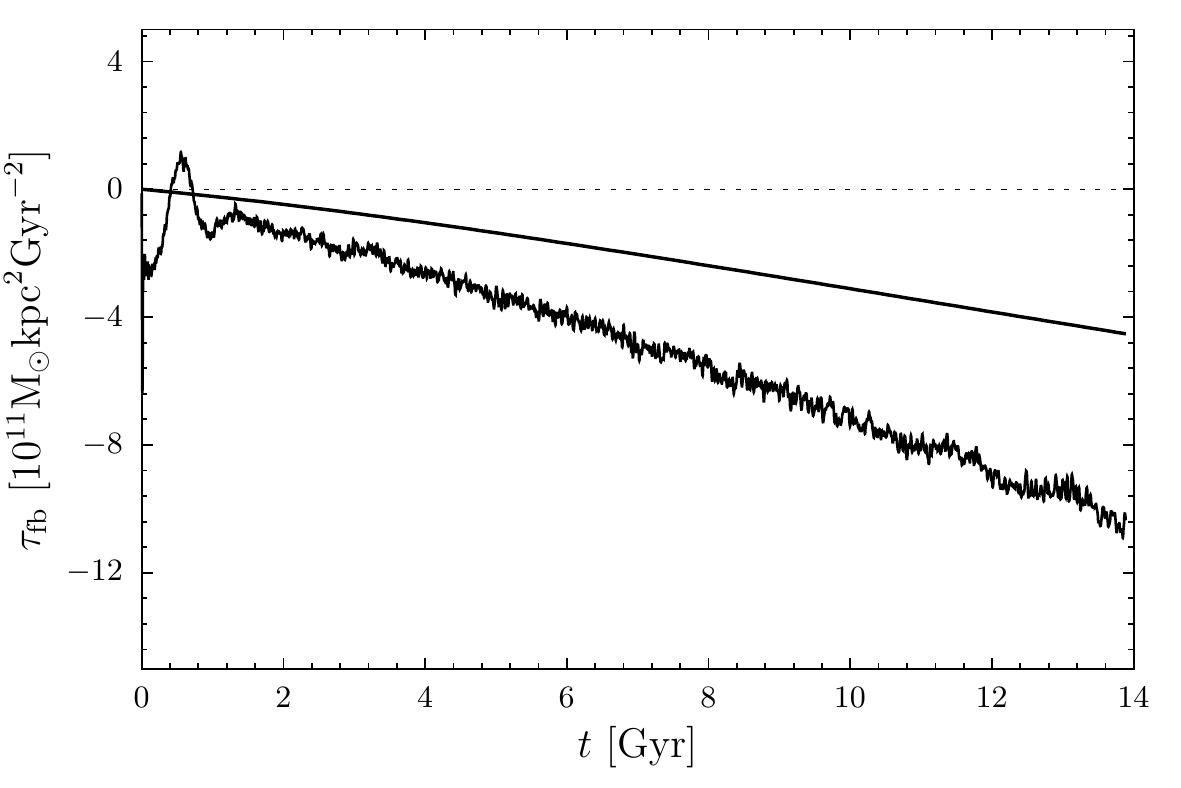}
    \caption{Comparison between dynamical feedback (smooth line) and total torque (jagged line) by the corotation resonance for $\eta=0.001$.}
    \label{fig:taufb_vs1Dsim}
  \end{center}
\end{figure}

Figure~\ref{fig:taufb_vs1Dsim} compares dynamical feedback with the total torque by the corotation resonance. The latter is computed by sampling particles from $f_0$ and evolving their slow variables forward in time. The total torque oscillates at early times as orbits captured at the time of bar formation collectively librate and phase mix \citep{Chiba2022Oscillating}. As the bar slows and grows in length, both the total torque and the dynamical feedback get stronger.

\begin{figure}
  \begin{center}
    \includegraphics[width=8.5cm]{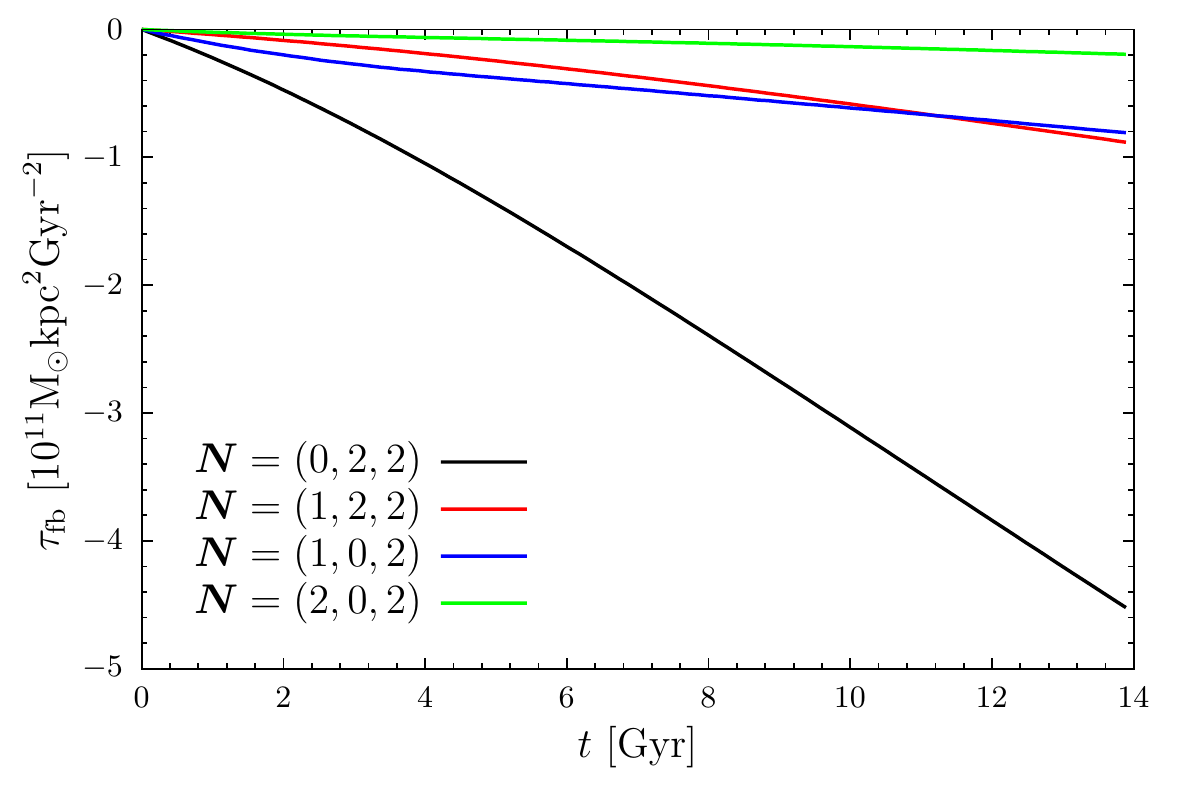}
    \caption{Dynamical feedback by the four strongest resonances for $\eta=0.001$.}
    \label{fig:taufb_N}
  \end{center}
\end{figure}

Figure~\ref{fig:taufb_N} compares dynamical feedback from different resonances. The corotation resonance is by far the dominant source of dynamical feedback, overwhelming others by more than a factor of 4.

\begin{figure}
  \begin{center}
    \includegraphics[width=8.5cm]{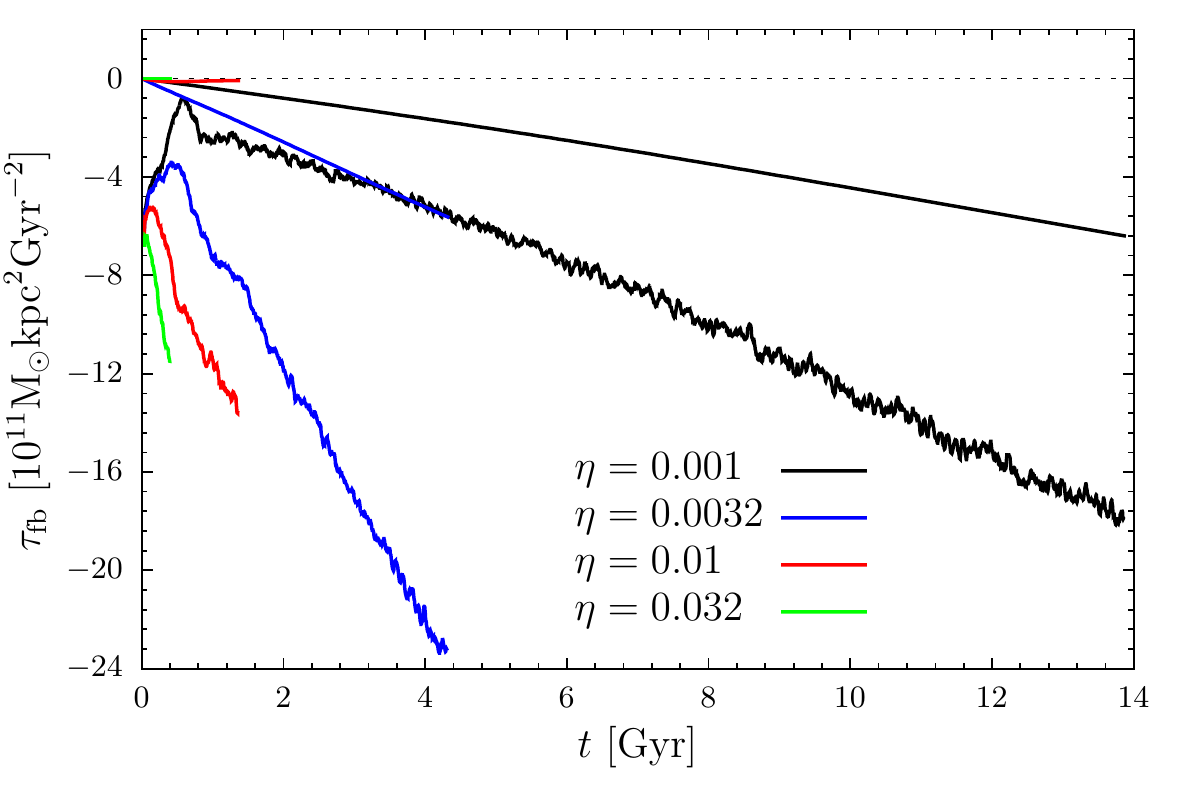}
    \caption{Comparison between the total dynamical feedback (smooth lines) and the total torque (jagged lines) computed from 3D test-particle simulations.}
    \label{fig:taufb_vs3Dsim}
  \end{center}
\end{figure}

Finally, Fig.~\ref{fig:taufb_vs3Dsim} compares the total dynamical feedback (sum of the four strongest resonances) with the total torque calculated from 3D test-particle simulations. The result suggests that, for small $\eta$ of order $10^{-3}$ (black and blue), around 30\% of the torque is composed of dynamical feedback. For large $\eta$ $(\gtrsim 10^{-2})$ in which the fast regime prevails (red and green), dynamical feedback is absent and most of the torque is due to dynamical friction.

\subsection{Dynamical feedback in infinite homogeneous system}
\label{subsec:dynamical_feedback_inf_homo_sys}

Dynamical feedback also applies to a point-mass perturber moving through an infinite homogeneous distribution of lighter particles. Here, trapped orbits correspond to those gravitationally bound to the moving perturber. When the perturber decelerates due to dynamical friction\footnote{The phase-space flow in a decelerating Kepler potential is studied by \cite{Namouni2007acceleratedKepler}}, it drags the trapped particles along with it. Hence, the trapped particles act to increase the effective inertia of the perturber. However, as described in the previous sections, conservation of phase-space volume requires that the deceleration of the trapped particles be accompanied by an acceleration of the surrounding untrapped particles, and it is the imbalance between the two that determines the net dynamical feedback. Let us consider the case in which the perturber is decelerating monotonically in its direction of motion along the $x$-axis, so $\vvel_{\rm p} = (v_{{\rm p}x},0,0)$, and the particles initially have an isotropic velocity distribution with a negative gradient $\pd f_0 / \pd v_x < 0$ for all $v_x > 0$.  Then, the phase-space density of the untrapped region at the current velocity of the perturber $v_{{\rm p}x}$, is larger than the phase-space density of the trapped region, which consists of orbits captured at $v_x > v_{{\rm p}x}$. Hence, the momentum gained by the untrapped orbits exceeds the momentum loss of the trapped orbits. Consequently, particles exert a net negative force (a positive feedback) on the perturber, meaning that the effective inertia of the perturber in fact \textit{decreases}. If the perturber accelerates instead, its effective inertia will increase. The conclusion is the same as in the bar-halo system. The classical formula of dynamical friction by \cite{Chandrasekhar1943DynamicalFriction} fails to capture this phenomenon because it is based upon the assumptions that the perturber has a constant velocity and that all field particles move pass the perturber on Keplerian hyperbolae, that is, all orbits are assumed to be unbound (untrapped).

\section{Discussion and conclusion}
\label{sec:conclusions}

We studied dynamical friction in the general fast-slow regime, where neither linear perturbation (fast limit) theory nor classical adiabatic (slow limit) theory is applicable. This intermediate regime is expected in realistic galaxy evolution, but has thus far eluded proper description.

We first showed that the phase-space dynamics around a drifting resonance is naturally described in the comoving frame of the resonance since the adiabatic condition there is better fulfilled than in the static frame. The phase-space flow in the comoving frame is determined by a single dimensionless parameter $s$, which is the ratio of the characteristic libration time to the resonance's migration time. We then solved the evolution of the distribution function numerically using the angle-averaged Hamiltonian, and provided an explanation to dynamical friction for a range of $s$, linking the theories developed earlier in the slow ($s=0$) and fast limits ($s\gg1$). The mechanism of dynamical friction can be summarized as follows:

(i) In the slow limit $(s = 0)$, the phase-space flow is similar to that of a pendulum, where trapped orbits librate inside the separatrix, while untrapped orbits circulate outside. As particles move along these orbits, they exchange angular momentum with the perturber. Dynamical friction arises from the imbalance between the gainers and losers, and is initially non-zero due to the negative gradient in the distribution function, but undergoes damped oscillations as the distribution phase mixes \citep{Chiba2022Oscillating,Banik2022Nonperturbative}.

(ii) In the general slow regime $(0 < s < 1)$, the trapped phase-space comove with the resonance, while its volume contracts. This contraction in the trapped volume allows the untrapped orbits to become open in phase space, transiting from one side of the resonance to the other. In this regime, in addition to the dynamical friction described above, the moving resonances give rise to \textit{dynamical feedback}, which refers to the transfer of angular momentum directly caused by the movement of the resonances. For example, a decelerating bar typically drives the resonances towards larger angular momentum $\Lz$, thus yielding $\Lz$ to the trapped orbits while receiving $\Lz$ from the untrapped orbits that leap over the drifting resonance. Due to the negative gradient in the distribution function, gainers of $\Lz$ generally outnumber the losers, resulting in a net negative torque on the perturber. This argument holds irrespective of the sign of the change in pattern speed $\dOmegap$. Dynamical feedback therefore facilitates the deceleration of the perturber (positive feedback) but hinders its acceleration (negative feedback). Since the phase-space volume diminishes in size with increasing $s$, dynamical feedback peaks at $s\sim 0.4$ and decays towards $s=1$.

(iii) In the general fast regime $(s > 1)$, the trapped region vanishes, and so does dynamical feedback. Here, angular momentum is transferred only through dynamical friction by the untrapped orbits. These untrapped orbits freely flow across the resonance and generate a striated perturbation downstream of the resonance. Since non-linear libration is absent in this regime, the perturbed distribution can be reproduced qualitatively by standard linear perturbation theory.

(iv) In the fast limit $(s \gg 1)$, the error of the linear solution diminishes since $s$ scales inversely with the perturbation amplitude. This is in line with the conclusion of \citetalias{Tremaine1984Dynamical}.

Dynamical feedback was first predicted by \citetalias{Tremaine1984Dynamical} although their description was incomplete as they only indicated feedback by the untrapped orbits. By identifying dynamical feedback as the torque arising from the phase-mixed distribution function, we derived the full analytic expression for dynamical feedback, accounting for both trapped and untrapped orbits. The untrapped component of our formula is shown to be equivalent to the equation derived in \citetalias{Tremaine1984Dynamical}. Our full formula highlights that dynamical feedback from each vicinity of the resonance is proportional to the difference between the mass of the trapped orbits and that of the untrapped orbits \textit{displaced by the trapped orbits}. This follows from the conservation of phase-space volume and is reminiscent of the Archimedes' principle in fluid dynamics -- just like the force required to lift an object submerged in a fluid, the torque required to shift the trapped island in phase space scales with the difference between the density of the trapped island and that of the background. Using our formula, we showed that dynamical feedback can comprise up to 30\% of the total torque on the Galactic bar, if the bar has been evolving with $\eta \sim 0.003$ as measured by \cite{Chiba2020ResonanceSweeping}.

Real galaxies are far more complex than the simple bar-halo system studied here: there are various sources of time-dependent perturbations (e.g. spiral arms, molecular clouds, orbiting substructures) that may stochastically scatter orbits and disrupt the resonant phase-space. The resulting diffusion in phase space is likely to sustain dynamical friction by recovering the gradient in the distribution function \citep{Hamilton2022BarResonanceWithDiffusion}, but is expected to suppress dynamical feedback as it acts to reduce the density difference between the trapped and untrapped regions. On the other hand, the self-gravity of the halo's response, which is neglected in this study, may coherently strengthen or weaken the bar's perturbation \cite[e.g.][]{Weinberg1989SelfGravitating,Chavanis2012Kinetic}. Linear theory predicts that self-gravity enhances dynamical friction on bars by almost a factor of two \cite[][]{Dootson2022}. This is in line with the halo's density response (Fig.~\ref{fig:dens_individual}) which is largely in phase with the bar in the fast-linear regime. In the slow-nonlinear regime, however, self-gravity may weaken both friction and feedback as the density response becomes out of phase with the bar. Work is in progress to quantify these effects.

\section*{Acknowledgements}

I thank J-B. Fouvry, N. Frankel, S. Hadden, C. Hamilton, J. Hunt, N. Murray, S. Tremaine and members of the NAOJ for valuable discussions. I am also grateful to my mentors R. Sch\"onrich, J. Magorrian and J. Binney for their helpful advice and encouragement. This work was supported by the Royal Society grant RGF$\backslash$R1$\backslash$180095 and the Natural Sciences and Engineering Research Council of Canada (NSERC), [funding reference \#DIS-2022-568580].

\section*{Data availability}

The codes used to produce the results are available from the corresponding author upon request.



\bibliographystyle{mnras}
\bibliography{references}



\appendix

\section{Numerical integration}
\label{sec:numerical_method}

In Sections \ref{sec:distribution_function} and \ref{sec:density_response}, we solve the collisionless Boltzmann equation (\ref{eq:CBE_sol_slow}) by numerical integration of the equations of motion averaged over the fast angles. The averaged Hamiltonian reads
\begin{align}
  \bH(\thetas,\vJ,t) = \bH_0(\vJ) + \Psi(\vJ,t) \cos \left(\thetas - \thetasres\right),
  \label{eq:app_H_averaged}
\end{align}
and the Hamilton's equations are
\begin{align}
  \dthetas &= \frac{\pd \bH}{\pd \Js} = \vN\!\cdot\!\vOmega(\vJ) - \Nphi \Omegap(t) + \frac{\pd \Psi(\vJ,t)}{\pd \Js} \cos \left(\thetas - \thetasres\right),
  \label{eq:app_EOM_dthetasdt} \\
  \dot{\Js} &= - \frac{\pd \bH}{\pd \thetas} = \Psi(\vJ,t) \sin \left(\thetas - \thetasres\right).
  \label{eq:app_EOM_dJsdt}
\end{align}
Since $\bH$ is non-separable, explicit symplectic integrators like leap-frog cannot be used. 
Following \cite{Weinberg2007BarHaloInteraction}, we employ the implicit midpoint rule, which is a second-order symplectic integrator \citep[][]{hairer2006geometric}. In practice, we tabulate the orbital frequencies $\vOmega(\Jr,L)$ and part of $\Psi$, namely the function $W_{lm}^{\Nr\Npsi}(\Jr,L,t)$ \citep[][Appendix B]{Tremaine1984Dynamical,Chiba2022Oscillating}. The tables are then interpolated linearly.

\section{The parameters $G$ and $\gamma$}
\label{sec:G_Gamma}

\begin{figure*}
  \begin{center}
    \includegraphics[width=17.5cm]{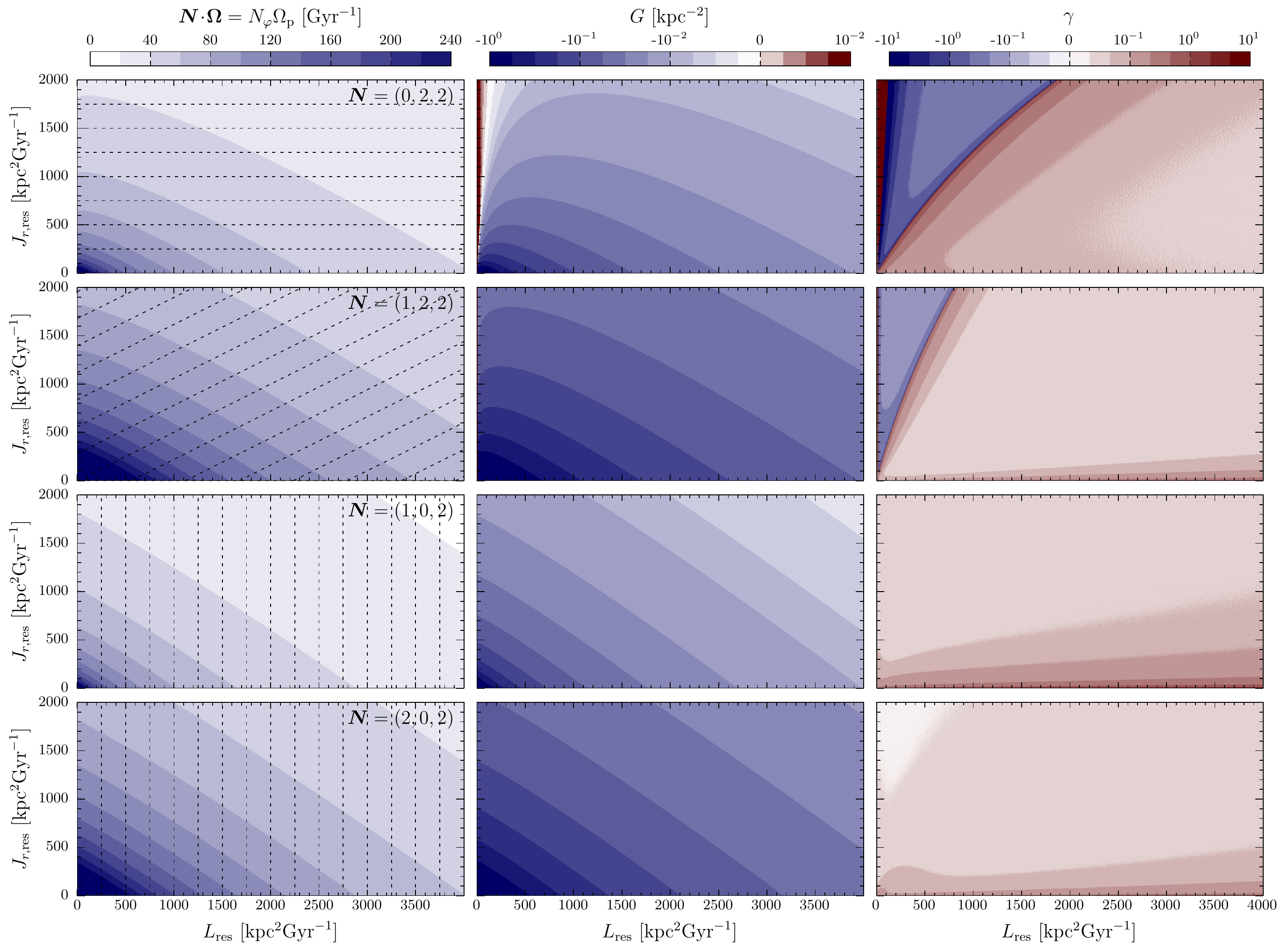}
    \caption{The resonant frequency $\vN \!\cdot\! \vOmega$ (left), its derivative $G=\pd(\vN \!\cdot\! \vOmega)/\pd \Js$ (middle), and the parameter $\gamma=\ddJres/\dJsres$ (right). The dotted lines mark contours of constant $\vJf$. The scales for $G$ and $\gamma$ are linear within $[10^{-2},10^2]$ and $[10^{-1},10^1]$, respectively, and logarithmic elsewhere.}
    \label{fig:NOmega_G_gamma}
  \end{center}
\end{figure*}

The left column of Fig.~\ref{fig:NOmega_G_gamma} plots the resonant frequency $\vN \!\cdot\! \vOmega|_\textrm{res} = \Nphi\Omegap$ for the major bar-halo resonances. Each contour represents the resonance curve at a given pattern speed. The dotted lines indicate the contours of constant fast actions $\vJf$.

The middle column of Fig.~\ref{fig:NOmega_G_gamma} plots $G \equiv \pd(\vN \!\cdot\! \vOmega)/\pd \Js|_\textrm{res}$ (equation \ref{eq:GF}), which is the derivative of the left column with respect to $\Js$ along constant $\vJf$ (dotted lines). $G$ is mostly negative apart from very small $L$ \citepalias[see also][Fig.~5]{Tremaine1984Dynamical}. Negative $G$ implies an increase in the position of resonance in $z$-angular momentum, $\Jsres(=L_{z,\textrm{res}}/\Nphi)$, with decreasing pattern speed $ \Nphi\Omegap = \vN \!\cdot\! \vOmega|_\textrm{res}$.

The right column of Fig.~\ref{fig:NOmega_G_gamma} plots the dimensionless parameter $\gamma = \ddJres/\dJsres$ (equation \ref{eq:gamma}), which measures how much the trapped volume changes as the resonance moves. Apart from regions where $G$ is close to zero, $\gamma$ is of order $10^{-1}$.


\bsp	
\label{lastpage}
\end{document}